%% file: manuscript.tex
\documentclass[10pt,journal,compsoc]{IEEEtran}
\input{includes.tex}
\input{colors.tex}

\input{macros.tex}
\input{spacing.tex}
\input{paper_shorthands.tex}

\begin{document}
\title{Whence to Learn? Transferring Knowledge\\ in Configurable Systems using \tool}

\author{Rahul Krishna,~%
    Vivek Nair,~%
    Pooyan Jamshidi,~%
    Tim Menzies,~\IEEEmembership{IEEE Fellow}

\IEEEcompsocitemizethanks{%
\IEEEcompsocthanksitem Rahul Krishna is with the Department
of Computer Science, Columbia University, New York, NY.
E-mail: \href{mailto:i.m.ralk@gmail.com}{i.m.ralk@gmail.com}.
\IEEEcompsocthanksitem Vivek Nair was with the Department
of Computer Science, North Carolina State University, Raleigh, NC.
E-mail: \href{mailto:vivekaxl@gmail.com}{vivekaxl@gmail.com}.
\IEEEcompsocthanksitem Tim Menzies is with the Department
of Computer Science, North Carolina State University, Raleigh, NC.
E-mail: \href{mailto:tim.menzies@gmail.com}{tim.menzies@gmail.com}
\IEEEcompsocthanksitem P. Jamshidi is with the Department of Computer Science and Engineering, University of South Carolina, Columbia, SC.
E-mail: \href{mailto:pooyan.jamshidi@gmail.com}{pooyan.jamshidi@gmail.com}
}
\thanks{Manuscript received December XX, 2019.}}

\input{body/0-frontmatter.tex}

\maketitle
\IEEEdisplaynontitleabstractindextext
\IEEEdisplaynontitleabstractindextext
\ifCLASSOPTIONcaptionsoff
 \newpage
\fi
 
\input{body/1-intro.tex}
\input{body/2-motivation.tex}

\input{body/3-problem_stmt.tex}

\input{body/4-beetle.tex}
\input{body/5-other_tl.tex}

\input{body/6-experimental_setup.tex}

\input{body/7-research_questions.tex}

\input{body/8-discussion.tex}

\input{body/9-threats.tex}
\input{body/10-related_work.tex}

\input{body/11-conclusion.tex}

\balance
\bibliographystyle{IEEEtran}
\bibliography{main} 
\input{body/12-bio.tex}

\end{document}

%% file: includes.tex
\usepackage{alltt}
\usepackage{multirow}
\usepackage{graphicx}
\usepackage{color}
\usepackage[labelfont=bf,textfont={bf}]{caption}
\usepackage{subcaption}
\usepackage{pifont}
\usepackage{boxedminipage}
\usepackage{notoccite}
\usepackage{array}
\usepackage{enumerate}
\usepackage{booktabs}
\usepackage[framemethod=tikz]{mdframed}
\usepackage{lipsum}
\usepackage{mathtools}
\usepackage{paralist}
\newcolumntype{C}[1]{>{\centering\let\newline\\\arraybackslash\hspace{0pt}}m{#1}}
\usepackage{graphics}
\usepackage{colortbl} 
\usepackage{subcaption}
\usepackage{blindtext, graphicx}
\usepackage{textcomp}
\usepackage{mathtools}
\usepackage[hidelinks]{hyperref}
\usepackage{amsmath}
\usepackage{amsfonts}
\usepackage{caption}
\usepackage{color}
\usepackage[final]{listings}
\usepackage{graphicx} 
\usepackage{multirow}
\usepackage{balance}
\usepackage{picture}
\usepackage{relsize}
\usepackage{multicol}
\usepackage{soul}
\usepackage{enumitem}
\usepackage{array}
\usepackage{makecell}
\usepackage{bigstrut}
\usepackage[pass]{geometry}
\usepackage[skins]{tcolorbox}
\usepackage{verbatim}
\usepackage{algorithm}
\usepackage{algorithmicx}
\usepackage{algpseudocode}
\usepackage[export]{adjustbox}
\usepackage{mathtools}
\usepackage{wrapfig}
\usepackage{fancyvrb}
\usepackage{soul}
\usepackage{color}
\usepackage{amsmath}
\usepackage{cite}
\newcommand{\subparagraph}{}
\usepackage{titlesec}
\usepackage[capitalise]{cleveref}

%% file: colors.tex
\sethlcolor{lightgray}
\definecolor{pink}{RGB}{252,145,149}
\definecolor{lightpink}{RGB}{252,145,149}
\definecolor{lightgray}{gray}{0.8}
\definecolor{darkgray}{gray}{0.6}
\definecolor{Gray}{rgb}{0.88,1,1}
\definecolor{Gray}{gray}{0.85}
\definecolor{Blue}{RGB}{0,29,193}
\definecolor{MyDarkBlue}{rgb}{0,0.08,0.45} 
\definecolor{pink}{RGB}{231,95,110}
\definecolor{greenish}{RGB}{182, 231, 142}
\definecolor{orangish}{RGB}{255, 206, 144}
\definecolor{lavender}{RGB}{225, 213, 231}
\definecolor{lightergray}{rgb}{0.85, 0.85, 0.85}
\definecolor{lightestgray}{rgb}{0.95, 0.95, 0.95}
\definecolor{codebg}{HTML}{F4F4F4}
\definecolor{blueish}{RGB}{177, 206, 232}
\definecolor{fig1_blue}{RGB}{85, 85, 255}
\definecolor{fig1_green}{RGB}{34, 148, 34}
\definecolor{gray05}{gray}{0.95}
\definecolor{gray08}{gray}{0.92}
\definecolor{gray10}{gray}{0.90}
\definecolor{gray12}{gray}{0.88}
\definecolor{gray15}{gray}{0.85}
\definecolor{gray18}{gray}{0.82}
\definecolor{gray20}{gray}{0.80}
\definecolor{gray25}{gray}{0.75}
\definecolor{gray30}{gray}{0.70}
\definecolor{gray35}{gray}{0.65}
\definecolor{gray40}{gray}{0.60}
\definecolor{gray45}{gray}{0.55}
\definecolor{gray50}{gray}{0.50}
\definecolor{gray55}{gray}{0.45}
\definecolor{gray60}{gray}{0.40}
\definecolor{gray65}{gray}{0.35}
\definecolor{gray70}{gray}{0.30}
\definecolor{gray75}{gray}{0.25}
\definecolor{gray80}{gray}{0.20}
\definecolor{gray85}{gray}{0.15}
\definecolor{gray90}{gray}{0.10}
\definecolor{gray95}{gray}{0.05}

%% file: macros.tex
\newcommand{\tion}[1]{\S\ref{sect:#1}}

\renewcommand{\footnotesize}{\scriptsize}

\newcommand{\bi}{\begin{itemize}}
\newcommand{\ei}{\end{itemize}}
\newcommand{\beq}{\begin{equation}}
\newcommand{\eeq}{\end{equation}}
\newcommand{\be}{\begin{enumerate}}
\newcommand{\ee}{\end{enumerate}}

\newcommand{\quart}[4]{\begin{picture}(80,6)
{\color{black}\put(#3,3){\circle*{4}}\put(#1,3){\line(1,0){#2}}}\end{picture}}
\newcommand{\quartex}[3]{
\begin{picture}(13,6)
  {
   \color{black}
    \put(#3,3)
    {\circle*{4}}
    \put(#1,3)
    {\line(1,0){#2}}
  }
\end{picture}
}
\tikzstyle{highlighter} = [lightgray, line width = \baselineskip]
\newcounter{highlight}[page]

\AtBeginShipout{\AtBeginShipoutUpperLeft{\ifthenelse{\value{highlight} > 0}{\tikz[remember picture, overlay]{\foreach \stroke in {1,...,\arabic{highlight}} \draw[highlighter] (begin highlight \stroke) -- (end highlight \stroke);}}{}}}

\newcommand{\etal}{{\em et al.\xspace}}

\newenvironment{steelblue}
{\color{black}}
{\color{black}}

\usetikzlibrary{shadows}
\newmdenv[
    tikzsetting= {fill=gray18},
    skipabove=0.33em,
    skipbelow=0.33em,
    linewidth=1pt,
    innerleftmargin=2pt,
    innerrightmargin=2pt,
    innertopmargin=2pt,
    innerbottommargin=1pt,
    linecolor=gray12,
    roundcorner=3pt, 
    shadow=false,
    shadowsize=5pt,
    shadowcolor=gray95
]{myshadowbox}

\newenvironment{result}
{\begin{myshadowbox} \textbf{Result:}\xspace}
{\end{myshadowbox}}

\newenvironment{goal}
{\vspace{0.15cm}
\noindent\begin{minipage}{\linewidth}
\begin{center}
\arrayrulecolor{black}
\begin{tabular}{|p{0.95\linewidth}|}
\hline\rowcolor{lightestgray}}
{\\\hline
\end{tabular}
\end{center}
\end{minipage}
\vspace{0.15cm}
}

\fvset{%
fontsize=\small,
numbers=left
}

%% file: spacing.tex
\setitemize{noitemsep,topsep=0pt,parsep=0pt,partopsep=0pt,leftmargin=*, wide=0pt}
\setenumerate{noitemsep,topsep=0pt,parsep=0pt,partopsep=0pt,leftmargin=*, wide=0pt}
\makeatletter
\g@addto@macro\normalsize{%
  \setlength\abovedisplayskip{2pt}
  \setlength\belowdisplayskip{2pt}
  \setlength\abovedisplayshortskip{2pt}
  \setlength\belowdisplayshortskip{2pt}
}
\makeatother
\titlespacing*{\section}{0pt}{0.3\baselineskip}{0.3\baselineskip}
\titlespacing*{\subsection}{0pt}{0.5\baselineskip}{0.5\baselineskip}
\titlespacing*{\subsubsection}{0pt}{0.1\baselineskip}{0.1\baselineskip}

%% file: paper_shorthands.tex
\newcommand{\measure}{$\mathit{NAR}$\xspace}
\newcommand{\tool}{BEETLE\xspace}

%% file: body/0-frontmatter.tex
\markboth{IEEE Transactions on Software Engineering, submitted May `19}{Krishna \MakeLowercase{\textit{\etal}}: Whence to Learn? Transferring Knowledge in Configurable Systems using \tool}

\IEEEtitleabstractindextext{%
\begin{abstract}
As software systems grow in complexity and the space of possible configurations increases exponentially, finding the near-optimal configuration of a software system becomes challenging. Recent approaches address this challenge by learning performance models based on a sample set of configurations. However, collecting enough sample configurations can be very expensive since each such sample requires configuring, compiling, and executing the entire system using a complex test suite. When learning on new data is too expensive, it is possible to use \textit{Transfer Learning} to ``transfer'' old lessons to the new context. Traditional transfer learning has a number of challenges, specifically, (a)~learning from excessive data takes excessive time, and (b)~the performance of the models built via transfer can deteriorate as a result of learning from a poor source. To resolve these problems, we propose a novel transfer learning framework called \tool, which is a ``bellwether''-based transfer learner that focuses on identifying and learning from the most relevant source from amongst the old data. This paper evaluates \tool with 57 different software configuration problems based on five software systems (a video encoder, an SAT solver, a SQL database, a high-performance C-compiler, and a streaming data analytics tool). In each of these cases, \tool found configurations that are as good as or better than those found by other state-of-the-art transfer learners while requiring only a fraction ($\frac{1}{7}$th) of the measurements needed by those other methods. Based on these results, we say that \tool is a new high-water mark in optimally configuring software.

\end{abstract}

}

%% file: body/1-intro.tex
\section{Introduction} 
A problem of increasing difficulty in modern software is finding the right set of {\em configurations} that can achieve the best performance. As more functionality is added to the code, it becomes increasingly difficult for users to understand all the options a software offers~\cite{xu2015hey,van2017automatic,JC:MASCOTS16,siegmund2012predicting,valov2015empirical,Siegmund:2015z,sarkar2015cost,oh2017finding,tang2018searching,nair2017faster,hsu2018micky,nair2017using}. It is hard to overstate the importance of good configuration choices and the impact poor choices can have. For example, it has been reported that for Apache Storm, the throughput achieved using the worst configuration was {\em 480 times slower} than that achieved by the best configuration~\cite{JC:MASCOTS16}.

Recent research has attempted to address this problem usually by creating accurate performance models that predict performance characteristics. While this approach is certainly cheaper and more effective than manual configuration it still incurs the expense of extensive data collection. This is undesirable, since the data collection must be repeated if the software is updated or the workload of the system changes. 

Rather than learning new configurations afresh, in this paper, we ask if we can learn from existing configurations. Formally, this is called ``transfer learning''; i.e., the transfer of information from selected
``\textit{source}'' software configurations running on one environment to learn a model for predicting the performance of some ``\textit{target}'' configurations in a different environment. Transfer learning has been extensively explored in other areas of software analytics~\cite{Nam2013, kocaguneli2011find,kocaguneli2012,turhan09,peters15, krishna18}. This is a practical possibility since often when a software is being deployed in a new environment
, there are examples of the system already executing under a different environment. To the best of our knowledge, this paper is among the earliest studies to apply transfer learning for performance optimization. Our proposed  method is significantly faster than any current state-of-the-art methods in identifying near-optimum configurations for a software system.

Transfer learning can only be useful in cases where the source environment is similar to the target environment. If the source and the target are not similar, knowledge should not be transferred. In such situations, transfer learning can be unsuccessful and can lead to a \textit{negative transfer}. Prior work on transfer learning focused on ``\textit{What to transfer}'' and ``\textit{How to transfer}'', by implicitly assuming that the source and target are related to each other. However, those work failed to address ``\textit{From where (whence) to transfer}''~\cite{pan2010survey}. Jamshidi \etal~\cite{jamshidi2017transfer2}  alluded to this and explained when transfer learning works but, did not provide a method which can help in selecting a suitable source. 

The issue of identifying a suitable source is a common problem in transfer learning. To address this, some researchers~\cite{krishna16,mensah17a, 
mensah17b, krishna18} have recently proposed the use of the \textit{bellwether} effect, which states that:

\noindent\begin{quote}
  ``When analyzing a community of software data, there is \underline{at least one} exemplary source data, called \underline{bellwether(s)}, which best defines predictors for all the remaining datasets \ldots''
\end{quote}
Inspired by the success of bellwethers in other areas, this paper defines and
evaluates a new transfer learner for software configuration called
\underline{Be}llw\underline{e}ther \underline{T}ransfer \underline{Le}arner (henceforth referred to as \textbf{\tool}). \tool can perform knowledge transfer using just a few samples from a carefully identified source environment(s). 

For evaluation, we explore five real-world software systems from different domains-- a video encoder, a SAT solver, a SQL database, a high-performance C-compiler, and a streaming data analytics tool (measured under 57 enviroments overall). In each case, we discovered that \tool found configurations as good as or better than those found by other state-of-the-art transfer learners while requiring only $\frac{1}{7}$-th of the measurements needed by those other methods. Reducing the number of measurements is an important consideration since collecting data in this domain can be computationally and monetarily expensive. 

Overall, this work makes the following contributions: 
\be
\item \textit{Source selection}: We show that the \textit{bellwether effect} exists in performance optimization and that we can use this to discover suitable sources (called bellwether environments) to perform transfer learning (see \S\ref{subsec:rq1}).
\item \textit{Cheap source selection:} 
\tool, using bellwethers, evaluates at most $\approx10\%$ of the configuration space (see \S\ref{sect:beetle}).
\item \textit{Simple Transfer learning using Bellwethers:} We develop a novel transfer learning algorithm using bellwether called \tool that exploits the bellwether environment to construct a simple transfer learner 
(see \S\ref{sect:beetle}).
\item \textit{More effective than non-transfer learning: } We show that using the \tool is \textit{better} than non-transfer learning approaches. It is also lot more economical (see \S\ref{sect:rq2}).
\item \textit{More effective than state-of-the-art methods: } Configurations discovered using the bellwether environment are better than the state-of-the-art 
methods~\cite{valov2017transferring, jamshidi2017transfer} (see \S\ref{subsec:rq4}).
\item \textit{Reproduction Package: } To assist other researchers, a reproduction package with all our scripts and data are available online (see \url{https://git.io/fjsky}).
\ee

The rest of this article is structured as follows: The remainder of this section presents the research questions asked here (\tion{rqs}) answered in this paper. \tion{motivation} presents some motivation for this work. \tion{formalization} describes the problem formulation and explains the concept of Bellwethers. \tion{beetle} describes
\tool followed by a quick overview of the prior work in transfer learning in performance configuration optimization in Section \tion{tl}.
In \tion{expt}, we present experimental setup and followed by answers to research questions in \tion{results}. In \tion{disc}, we discuss our findings further and answer some additional questions pertaining to our results. \tion{threats} discusses some threats to validity, related work and conclusion are presented in \tion{related} and \tion{conclusion} respectively.

\input{intro_figure.tex}

\subsection{Research questions}
\label{sect:rqs}
\begin{itemize}
    \item[\textbf{RQ1:}]\textbf{Does there exist a Bellwether Environment?}
    First, we ask if there exist bellwether environments to train transfer learners for performance optimization. We hypothesize that, if these bellwether environments exist, we can improve the efficacy of transfer learning.

    \begin{result}
        We find that bellwether environments are prevalent in performance optimization. That is, in each of the software systems, there exists at least one environment that can be used to construct superior transfer learners.
    \end{result}
        
    \item[\textbf{RQ2:}]\textbf{How many performance measurements are required to discover bellwether environments?}
    Having established that bellwether environments are prevalent, the purpose of this research question is to establish how many performance measurements are needed in each of the environments to discover these bellwether environments.

    \begin{result}
        We can discover a potential bellwether environment by measuring as little as 10\% of the total configurations across all the software system.
    \end{result}

    \item[\textbf{RQ3:}]\textbf{How does \tool compare with other non-transfer-learning based methods?}
    The alternative to transfer learning is just to use the target data to find the near-optimal configurations. In the literature are many examples of this ``non-transfer'' approach~\cite{guo2013variability, sarkar2015cost, nair2017using, nair2017faster} and for our comparisons, we used the current state-of-the-art performance optimization model proposed by Nair \etal~\cite{nair2017faster}. 

    \begin{result}
        Our experiments demonstrate that transfer learning using bellwethers (\tool) outperforms other methods that do not use transfer learning both in terms of cost and the quality of the model.
    \end{result}

    \item[\textbf{RQ4:}]\textbf{How does \tool compare to state-of-the-art transfer learners?}
    The final research question compares \tool with two other state-of-the-art transfer learners used commonly in performance optimization (for details see \tion{tl}). The purpose of this research question is to determine if a simple transfer learner like \tool with carefully selected source environments can perform as well as other complex transfer learners  that do not perform any source selection.    

    \begin{result}
        We show that a simple transfer learning using bellwether environment (\tool) just as good as (or better than) current state-of-the-art transfer learners.
    \end{result}
\end{itemize}

%% file: intro_figure.tex
\begin{figure}[t!]
    \centering
    \includegraphics[width=\linewidth]{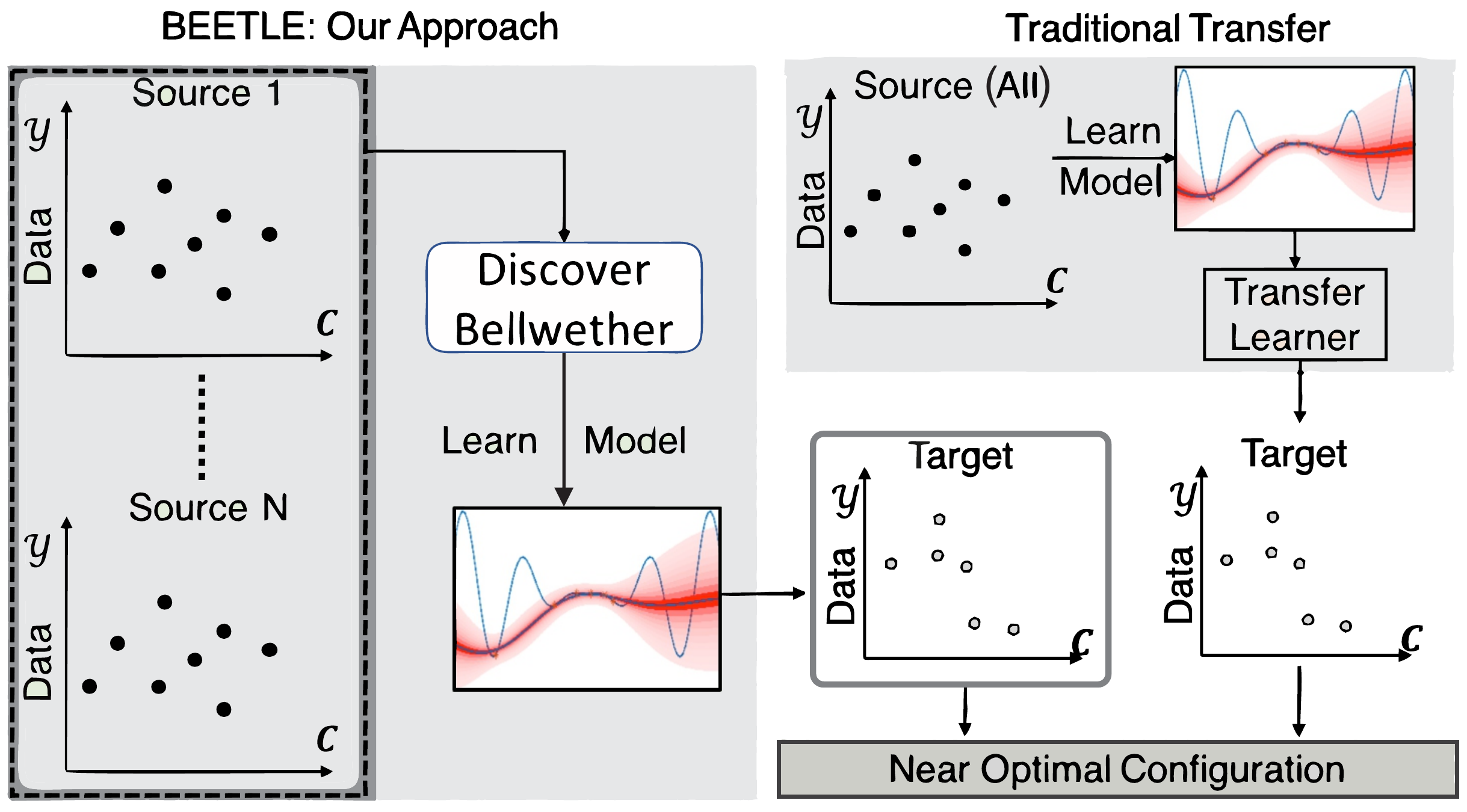}
    \caption{Traditional Transfer Learning compared with using bellwethers to 
    discover near optimal configurations.}
    \label{fig:intro_fig}
\end{figure}

%% file: body/2-motivation.tex
\section{Motivation}
\label{sect:motivation}
With the appearance of continuous software engineering and devops,  {\em configurability} has become a primary concern of software engineers. System administrators today develop and use different versions software programs under running several different workloads and in numerous environments. In doing so, they try to apply software engineering methods to best configure these software systems. Despite their best efforts, the available evidence is that they need to be better assited in making all the configuration decisions. Xu \etal~\cite{xu2015hey} reports that, when left to their own judgementments, developers ignore up to 80\% of configuration options, which exposes them to many potential problems. For this reason, the research community is devoting a lot of effort to configuration studies, as witnessed by many recent software engineering research publications~\cite{siegmund2012predicting, valov2015empirical, Siegmund:2015z, sarkar2015cost, oh2017finding,nair2017using, tang2018searching, nair2017faster, nair2018finding}. For details, see \tion{related} for the additional related work.

Without automatic support (e.g., with systems like \tool), humans find it difficult to settle on their initial choice for software configurations. The available evidence~\cite{van2017automatic,JC:MASCOTS16, herodotou2011starfish} shows that system administrators frequently make poor configuration choices. Typically, off-the-shelf defaults are used, which often behave poorly. There are various examples presented in the literature which have established that choosing default configuration can lead to sub-optimal performance. For instance, Van Aken \etal report that the default MySQL configurations in 2016 assume that it will be installed on a machine that has 160MB of RAM (which, at that time, was incorrect by, at least, an order of magnitude)~\cite{van2017automatic}. Also, Herodotou \etal~\cite{herodotou2011starfish} report that default settings for Hadoop results in the \textit{worst possible} performance. 

Traditional approaches to finding good configuration are very resource intensive. A typical approach uses sensitivity analysis~\cite{saltelli2000sensitivity}, where performance models are learned by measuring the performance of the system under a limited number of sampled configurations. While this approach is cheaper and more effective than manual exploration, it still incurs the expense of extensive data collection about the software~\cite{guo2013variability, sarkar2015cost, siegmund2012predicting, nair2017faster, nair2017using, nair2018finding, oh2017finding, guo2017data,JC:MASCOTS16}. This is undesirable since this data collection has to be repeated if ever the software is updated or the environment of the system changes abruptly. While we cannot tame the pace of change in modern software systems, we can reduce the data collection effort required to react to that change. The experiments of this paper make the case that  \tool scales to some large configuration problems, better than the prior state of the art. Further, it does so using fewer measurements than existing state-of-the-art methods.

Further, we note that \tool is particularly recommended in highly dynamic projects where the environments keep changing. When context changes, so to must the solutions applied by software engineers. When frequently re-computing best configurations, it becomes vitally important that computation cost is kept to a minimum. Amongst the space of known configuration tools, we most endorse \tool for very dynamic environments. We say this since, of all the systems surveyed here, \tool has the lowest CPU cost (and we conjecture that this is so since \tool makes the best use of old configurations).

As a more concerete example, consider an organization that runs, say, $N$ heavy Apache Spark workloads on the cloud. To optimize the performance of Apache Spark on the given workloads, the DevOps Team need to find the optimal solutions for each of these workloads, i.e., conduct performance optimization $N$ times. This setup has two major shortcomings: {\em hardware change} and {\em workload change}.

 \textit{Hardware Change}: Even though the DevOps engineer of a software system performs a performance optimization for a specific workload in its staging environment, as soon as the software is moved to the production environment the optimal configuration found previously may be inaccurate. This problem is further accentuated if the production environment changes due to the ever-expanding cloud portfolios. It has been reported that cloud providers expand their cloud portfolio more than 20 times in a year~\cite{ec2history}.

\textit{Workload Change}: The developers of a database system can optimize the system for a read-heavy workload, however, the optimal configuration may change once the workload changes to, say, a write-heavy counterpart. The reason is that if the workload changes, different functionalities of the software might get activated more often and so the nonfunctional behavior changes too. This means that as soon as a new workload is introduced (new feature in the organization\textquotesingle s product) or if the workload changes, the process of performance optimization needs to be repeated. 

Given the fragility of traditional performance optimization, it is imperative that we develop a method to learn from our \textit{previous experiences} and hence reduce the burden of having to find optimum configurations ad nauseam.

%% file: body/3-problem_stmt.tex
\section{Definitions and Problem Statement}
\label{sect:formalization}

\noindent\textbf{Configuration: }A software system, $\mathcal{S}$, may offer a number of configuration options that can be changed. We denote the total number of configuration options of a software system $\mathcal{S}$ as $N$. A configuration option of the software system can either be a (1)~continuous numeric value or a (2)~categorical value. This distinction is very important since it impacts the choice of machine learning algorithms. The configuration options in all software systems studied in this paper are a combination of both {\em categorical} and {\em continuous} in nature. The learning algorithm used in this paper namely, \textit{Regression Trees}, are particularly well suited to handle such a combination of continuous and categorical data.

A configuration is represented by $c_{i}$, where $i$ represents the $i^{th}$ configuration of a system. A set of all configurations is called the \textit{configuration space}, denoted as $\mathcal{C}$. Formally, $\mathcal{C}$ is a Cartesian product of all possible options $\mathcal{C}$ = Dom($c_1$) $\times$ Dom($c_2$) $\times$ ... $\times$ Dom($c_N$), where $\text{Dom}(c_i)$ is either $\mathbb{R}$ (Real Numbers) or $\mathbb{B}$ (Catergorical/Boolean value) and $N$ is the number of configuration options. 
\begin{figure}[h!]
\setlength{\belowcaptionskip}{-0.5\baselineskip}
\resizebox{.99\linewidth}{!}{
\begin{tabular}{l|l|l|l|r}
           & \texttt{ATOMIC} & \texttt{USE\_LFS} & \texttt{SECURE} & LATENCY ($\mu s$) \\\hline
$c_1$        & 0              & 0              & 0               & 100              \\\hline
$c_2$        & 0              & 0              & 1              & 150              \\\hline
\vdots & \vdots     & \vdots     & \vdots     & \vdots     \\\hline
$c_N$        & 1              & 1              & 1              & 400              
\end{tabular}}
  \caption{Some configuration options for \texttt{SQLite}.}
  \label{fig:sample_config}
\end{figure}

As a simple example, consider a subset of configuration options from \texttt{SQLite}, i.e., $\mathcal{S}\equiv\text{\texttt{SQLite}}$. This is shown in \cref{fig:sample_config}. The subset of \texttt{SQLite} offers three configuration options namely, \texttt{ATOMIC} (atomic delete), \texttt{USE\_LFS}  (use large file storage), and \texttt{SECURE} (secure delete), ie., $N=3$. The last column contains the \textit{latency} in $\mu s$ when various combinations of these options are chosen.

\noindent\textbf{Environment: }
As defined by Jamshidi \etal \cite{jamshidi2017transfer2}, the different ways a software system is deployed and used is called its {\em environment} ($e$). The environment is usually defined in terms of: (1) \textbf{\textit{workload}} ($w$): the input which the system operates upon; (2) \textbf{\textit{hardware}} ($h$): the hardware on which the system is running; and (3) \textbf{\textit{version}} ($v$): the state of the software. 

Note that, other environmental changes might be possible (e.g., JVM version used, etc.). For example, consider software system Apache Storm, here
we must ensure that an appropriate JVM is installed in an environment before it can be deployed in that environment. Indeed, the selection of one version of a JVM over another can have a profound performance impact. However, the perceived improvement in the performance is due to the optimizations in JVM, not the original software system being studied. Therefore, in this paper, we do not alter these other factors which do not have a direct impact on the performance of the software system. 

\noindent The following criteria is used to define an environment:
\be
  \item Environmental factors of the software systems that we can vary in the deployment stack of the system. This prevents us from varying factors such as the JVM version, CPU frequency, system software, etc., which define the deployment stack and not the software system. 
  \item Common changes developers choose to alter in the software system. In practice, it is these factors that affect the performance of systems the most~\cite{jamshidi2017transfer2, jamshidi2017transfer, valov2017transferring, valov2015}. 
  \item Factors that are most amenable for transfer learning. Preliminary studies have shown that factors such as workload, hardware, and software version lend themselves very well to transfer learning~\cite{jamshidi2017transfer2, jamshidi2017transfer}.
\ee
For a more detailed description of the factors that were changed and those that were left unchanged, see~\cref{tab:datasets}.

Formally, we say an environment is  $\mathit{\mathbf{e}}=\{w, h, v\}$ where $w \subseteq W$, $h \subseteq H$, and $v \subseteq V$. Here, $W, H, V$ are the space of all possible hardware changes $H$;  all possible software versions $V$, and all possible workload changes $W$. With this, the environment space is defined as $\mathcal{E}\subset\{W\times H\times V\}$, i.e., a subset of environmental conditions $\mathit{\mathbf{e}}$ for various workloads, hardware, and environments.

\noindent\textbf{Performance: } 
For each enviroment $\mathit{\mathbf{e}}$, the instances in our data are of the form $\{(c_1, y_1), ..., (c_N,\; y_N)\}$, where $c_i$ is a vector of configurations of the i-th example and it has a corresponding performance measure $y_i \in Y_{S,c,e}$ associated with it. We denote the performance 
measure associated with a given configuration ($c_i$) by $y=f(c^i)$. We 
consider the problem of finding the near-optimal configurations ($c^{*}$) such 
that $f(c^{*})$ is better than other configurations in $C_{A,e}$, i.e.,
$$
\centering
   {\begin{array}{*{20}{l}}
   \centering
{f(c^{*}) \le f(c){\rm{~}}\forall c \in {C_{A,h,w,v}}\setminus c^{*} }&{{\text {for min objective}}}\\
{f(c^{*}) \ge f(c){\rm{~}}\forall c \in {C_{A,h,w,v}}\setminus c^{*} }&{{\text {for max objective}}}
\end{array}}
$$
\noindent\textbf{Bellwethers:}
In the context of performance optimization, the bellwether effect states that: \textit{For a configurable system, when performance measurements are made 
  under 
  different environments, then
   among those environments there exists one exemplary environment, called 
  the bellwether, which can be used determine near optimum configuration for 
  other environments for that system}.
We show that, when performing transfer 
learning, there are exemplar source environments called the bellwether 
environment(s) ($\mathcal{B}={e_{s1}, e_{s2},...,e_{sn}}\subset E$), which 
are the best source environment(s) to find near-optimal configuration for the 
rest of the environments ($\forall e \in E\setminus \mathcal{B}$).

\noindent\textbf{Problem Statement}: The problem statement of this paper:

\begin{goal}
Find a near-optimal configuration for a target environment ($S_{e_t}$), by learning from the measurements ($\langle c,y\rangle$) for the same system operating in different source environments ($S_{e_s}$).
\end{goal}

In other words, we aim to reuse the measurements from a system operating in an environment to optimize the same system operating in the different environment thereby reducing the number of measurements required to find the near-optimal configuration.

%% file: body/4-beetle.tex
\section{\tool: \underline{Be}llweth\underline{e}r \underline{T}ransfer \underline{Le}arner}
\label{sect:beetle}

\input{approach.tex}

This section describes \tool, a bellwether based approach that finds the near-optimal configuration using the knowledge in the ``bellwether'' environment. \tool can be separated into two main steps: (i) {\em Discovery:}~finding the bellwether environment, and (ii) {\em Transfer:} using the bellwether environment to find the near-optimal configuration for target environments. These steps will are explained in greater detail in \tion{finding} and \tion{transferring}. We outline it below,

\be
\item \textit{Discovery:} Leverages the existence of the bellwether effect to {\em discover} which of the available environments are best suited to be a {\em source enviroment} (known as the \textit{bellwether environment}). To do this, \tool uses a {\bf racing algorithm} to sequentially evaluate candidate environments~\cite{birattari2002racing}. In short,

\begin{enumerate}[leftmargin=1em]
    \item A fraction (about 10\%) of all available data is sampled. A prediction model is built with these sampled datasets.
    \item Each enviroment is used as a {\em source} to build a prediction model and all the others are used as {\em targets} in a round-robin fashion.
    \item Performance of all the enviroments are measured and are statistically ranked from the best source environemnt to the worst. Environments with a {\em poor} performance (i.e., those ranked last) are eliminated.
    \item For the remaining enviroments, another 10\% of the samples are added and the steps (a)--(c) are repeated.
    \item When the ranking order doesn't change for a fixed number of repeats, we terminate the process and nominate the best ranked enviroment(s) as the bellwether.
\end{enumerate}

  \item \textit{Transfer:} Next, to perform transfer learning, we just use these bellwether environments to train a performance prediction model with  {\em regression trees}~\cite{breiman2017classification}.
\ee
We conjecture that once a \textit{bellwether source environment} is identified, it is possible to build a simple transfer model without any complex methods and still be able to discover near-optimal configurations in a target environment. 

\subsection{{\em Discovery}: Finding Bellwether Environments}
\label{sect:finding}

In the previous work on bellwethers~\cite{krishna18}, the discovery process involved a round-robin experimentation comprised of the following steps:
\be
\item Pick an enviroment $\mathit{\mathbf{e}}_i$ from the space of all available enviroments, i.e., $\mathit{\mathbf{e}}_i \in \mathcal{E}$.  
\item Use $\mathit{\mathbf{e}}_j$ as a {\em source} to build a prediction model.
\item Using all the {\em other} enviroments $\mathit{\mathbf{e}}_j \in \mathcal{E}$ and $\mathit{\mathbf{e}}_j \neq \mathit{\mathbf{e}}_i$ as the {\em target}, determine the prediction performance of $\mathit{\mathbf{e}}_i$.
\item Next, repeat the steps by choosing a different $\mathit{\mathbf{e}}_i \in \mathcal{E}$
\item Finally, rank the performances of all the enviroments and pick the best ranked enviroment(s) as bellwether(s).
\ee

The above methodology is a form of an exhaustive search. While it worked for the relatively small datasets in \cite{krishna16, krishna18}, the amount of data in this paper is sufficiently large (see~\cref{tab:datasets}) that scoring all candidates using every sample is too costly. More formally, let us say that we have $M$ candidate enviroments with $N$ measurements each. The classical approach, described above, will construct $M$ models. If we assume that the model construction time is a function of number of samples $f(N)$, then for one round in the round-robin, the computation time will $O(M\cdot f(N))$. Since this is repeated $M$ times for each enviroment, the total computational complexity is $O(M^2\cdot f(n))$. When $M$ and/or $N$ is/are extremely large, it becomes necessary to seek alternative methods. Therefore, in this paper, we use a racing algorithm to achieve computational speedups.

Instead of evaluating every available instance to determine the best source enviroment, {\em Racing algorithms} take the following steps:
\begin{itemize}
    \item Sample a small fraction of instances from the original enviroments to minimize computational costs. 
    \item Evaluate the performance of enviroments statistically.
    \item Discarded the enviroments with the poorest performance.
    \item Repeated the process with the remaining datasets with slightly larger sample size. 
\end{itemize}

\Cref{fig:approach_b}(a)
shows how \tool
 eliminates  inferior environments at every iteration (thus
 reducing the overall  number of environments evaluated). Since  each iteration  only uses a small sample of the available data, the model building time also reduces significantly. It has been shown that racing algorithms are extremely effective in model selection when the size of the data is arbitrarily large~\cite{birattari2002racing, loh2013faster}.

In \Cref{fig:approach_b}(b), we illustrate the discovery of the bellwether environments with an example. Here, there are two groups of environments:
\begin{enumerate}[leftmargin=1.5em, label=(\roman*)]
    \item Group 1: Environments $e_1, e_2,..., e_7$, for which performance measurements have been gathered. One or more these environment(s) are potentially bellwether(s). 
    \item Group 2: Environments $e_8, e_9,..., e_{12}$, these represent the target environments, for which need to determine an optimal configuration. 
\end{enumerate}

In the discovery process, \tool's objective is to {\em find bellwethers} from among the environments in Group 1. And, later in the {\em Transfer} phase, we use the bellwether enviroments to find the near-optimal configuration for the target environments from Group 2.Note that, for the enviroments in Group 2, we \textit{do have to make any measurements} regarding it's performance. Having found bellwether enviroment(s) from Group 1, it is sufficient to just use the bellwether enviroment(s) to predict the optimal configurations for the enviroments in Group 2.

\Cref{fig:approach_b}(d) outlines a pseudocode for the algorithm used to find bellwethers. The key steps are listed below:
\bi 
\item {\em Lines 3--5}: Randomly sample a small subset of configurations from the source environments. The size of the subset (of configurations) is controlled by a predefined parameter \textit{frac}, which defines the percent of configurations to be sampled in each iteration. 

\item {\em Line 6--7}: Calculate sampling cost for the configurations. 

\item {\em Line 8--9}: Use the sampled configurations from each environment as a {\em source} build a prediction model with regression trees. For all the remaining enviroments, this regression tree model is used to predict for optimum configuration. After using every enviroment as a {\em source}, the environments are ranked from best to worst using the evaluation criteria discussed in \tion{nar}.
\item {\em Line 10--14}: We check to see if the rankings of the enviroments have changed since the last iteration. If not, then a ``life'' is lost. We go back to {\em Line 3} and repeat the process. When all lives are expired, or we run out of the budget, the search process terminates. This acts as an early stopping criteria, we need not sample more data if those samples do not help in improving the outcome.

\item {\em Line 15--17}: If there is some change in the rankings, then new configuration samples are informative and the environments that are \textit{ranked last} are eliminated. These environments are not able to find near-optimal configurations for the other environments and therefore cannot be bellwethers. 

\item {\em Line 18}: Once we have exhausted all the lives or the sampling budget, we simply return the source project with the best rank. These would be the bellwether enviroments.
\ei 

 {\color{black} On {\em line 6--7} we measure the sampling cost. In our case, we use the number of samples as a proxy for cost. This is because each measurement consumes computational resources, which in turn has a monetary cost. Therefore, it is a commonplace to set a budget and sample such that the budget is honored. Our choice of using the number of measurements as a cost measure was an engineering judgment; this can be replaced by any user-defined cost function such as (1)~the actual cost, or (2)~the wallclock time. The accuracy of either of the above is dependent on the business context. If one is either constrained by the runtime or there is a large variance in the measurements of time per configuration, then the wallclock time might be a more reasonable measure. On the other hand, if the cost of measurements is the limiting factor, it makes sense to use the actual measurement cost. Using the number of samples encompasses these two factors since it both costs more money and takes time to obtain more samples.} In the ideal case, we would like to have performance measurements for all possible configurations of a software system. But this is not practical because certain systems have over $2^{50}$ unique configurations (see \cref{tab:datasets}).

It is entirely possible for the \textit{FindBellwether} method to identify multiple bellwethers (e.g., in the case of \Cref{fig:approach_b}(b) the bellwethers were $e_1$ and $e_2$). When mutliple bellwethers are found, we may use (a) any one of the bellwether enviroments at random, (b) use all the enviroments, or (c) use heuristics based on human intuition. In this paper, we pick one enviroment from among the bellwethers at random. As long as the chosen project is among the bellwether enviroments, the results remain unchanged.

The \tool approach assumes that a \textit{fixed} set of enviroments exist from which we pick one or more bellwethers. But, approach would work just as well where new measurements from new enviroments are added.
Specifically, when more environments are added into a project, it is possible that the newly added environment could be the bellwether. Therefore, we recommend repeating \textit{FindBellwether} method prior to using the new enviroment. Note that, repeating \textit{FindBellwether} for new enviroments would add minimal computational overhead since the measurements have already been made for the new enviroments. Also note that, this approach of revisiting \textit{FindBellwether} on availability of new data, has been previously been proposed in other domains in software engineering~\cite{krishna16,krishna18}.

\subsection{{\em Transfer}: Using the Bellwether Environments}
\label{sect:transferring}
Once the bellwether environment is identified, it can be used to find the near-optimal configurations of target environments. As shown in \Cref{fig:approach_b}(c), \textit{FindBellwether} eliminates enviroments that are not potentially bellwethers and returns only the candidate bellwether enviroments.
For the remaining target enviroments, we use the model built with the bellwether enviroments to identify the near optimal configurations.

\Cref{fig:approach_b}(e) outlines the pseudocode used to perform the transfer. The key steps are listed below:
\bi
\item {\em Line 9-10}: We use the {\em FindBellwether} from \Cref{fig:approach_b}(d) to identify the bellwether enviroments. 
\item {\em Line 13-14}: If there exists more than one bellwether, we randomly chose one among them be used as the bellwether enviroment.
\item {\em Line 15-16}: The configurations from the bellwether and their corresponding performance measures are used to build a prediction model using regression trees.
\item {\em Line 17-18}: Predict the performances of various configurations from the target enviroment. 
\item {\em Line 19-20}: Return the best configuration for the target. 
\ei

Note that, on {\em Line 10}, we use \textit{regression trees} to make predictions. It has been the most preferred prediction algorithm in this domain~\cite{guo2013variability, sarkar2015cost, nair2017faster}. This is primarily because much of the data used in this domain are a mixture numerical and categorical attributes. Given configuration measurement in the form $\{(c_i, y_i)\}$, $c_i$ is a vector of categorical/numeric values and $y_i$ is a continuous numeric value. For such data, regression trees are the best suited prediction algorithms~\cite{nair2017using, nair2018finding, guo2017data, valov2017transferring}. 

In terms of computational complexity in comparison with previous methods \cite{krishna16, krishna18}, BEETLE offers noticible speedups. Given that we have $\mathit{M}$ enviroments and $\mathit{n}$ measurements in each enviroment, we may categorize the speedups into the following cases:
\bi
\item {\em Best Case:} Here, we expect the racing algorithm of \tool to eliminate atleast half of the non-bellwether enviroments at every iteration. This gives us a recurrence relation of $T(M) = T(M/2) + f(n)$, which gives us a best case complexity of $\mathit{O(log_2(M)\cdot f(n))}$. 
\item {\em Worst Case:} Here, we expect the racing algorithm of \tool to eliminate \textit{just one} non-bellwether enviroment at every iteration. This gives us a recurrence relation of $T(M) = T(M-1) + f(n)$, which gives us a worst case complexity of $\mathit{O(M^2 \cdot f(n))}$. Note that this worst case is same as the complexity of \cite{krishna16, krishna18}.  
\ei
In practice, we note that the average case speedup is somewhere in between that of the best case (i.e., $\mathit{O(log_2(M)\cdot f(n))}$) and the worst case (i.e., $\mathit{O(M^2 \cdot f(n))}$).

%% file: approach.tex
\begin{figure*}[t]
    \centering
    \includegraphics[width=\linewidth]{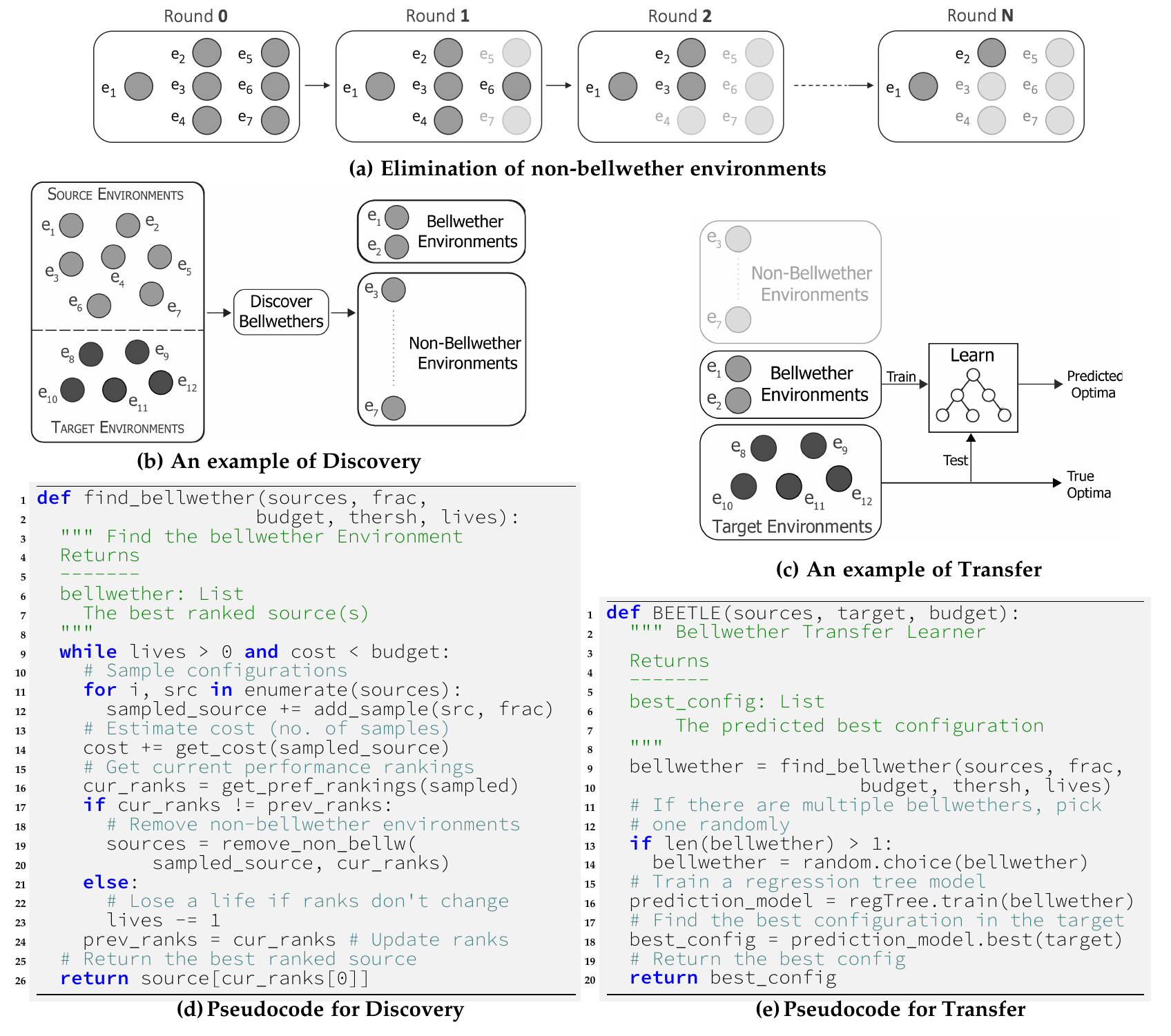}
    \caption{\tool framework and Pseudocode.}
    \label{fig:approach_b}
    \end{figure*}

%% file: body/5-other_tl.tex
\section{Other Transfer Learning Methods}
\label{sect:tl}

This section describes the methods we use to compare \tool against.
These alternatives are  (a)~two state-of-the-art transfer learners for performance optimization: Valov \etal~\cite{valov2017transferring} and Jamshidi \etal~\cite{jamshidi2017transfer}; and (b) a non-transfer learner: Nair \etal~\cite{nair2017using}. 
\input{other_pseudo.tex}
\subsection{Transfer Learning with Linear Regression}

Valov \etal~\cite{valov2017transferring} proposed an approach for transferring performance models of software systems across platforms with \textit{different hardware settings}. The method consists of the following two components: 
\bi
\item \textit{Performance prediction model:} The configurations on a source hardware are sampled using \textit{Sobol} sampling. The number of configurations is given by $T\times N_f$, where $T={3, 4, 5}$ is the \textit{training coefficient} and $N_f$ is the number of configuration options. These configurations are used to construct a \textit{Regression Tree} model.
\item \textit{Transfer Model:} To transfer the predictions from the source to the target, a linear regression model is used since it was found to provide good approximations of the transfer function.  To construct this model, a small number of random configurations are obtained from {\em both the source and the target}. Note that this is a shortcoming since, without making some preliminary measurements on the target, one cannot begin to perform transfer learning.
\ei

\subsection{Transfer Learning with Gaussian Process} 

Jamshidi \etal~\cite{jamshidi2017transfer} took a slightly different approach to transfer learning. They used Multi-Task Gaussian Processes (GP) to find the relatedness between the performance measures in source and the target. The relationships between input configurations were
captured in the GP model using a covariance matrix that
defined the kernel function to construct the Gaussian processes model. To encode the relationships between the measured performance of the source and the target, a scaling factor is used with the above kernel. The new kernel function is defined as follows:
\begin{equation}
  \setlength{\belowdisplayskip}{1em}
  \setlength{\abovedisplayskip}{1em}
  k(s, t, f(s), f(t)) = k_t(s, t) \times k_{xx}(f(s), f(t)),
\end{equation}\label{eq:gpkernel}
where $k_t(s,t)$ represents the multiplicative scaling factor. $k_t(s,t)$ is given by the correlation between source f(s) and target f(t) function, while $k_{xx}$ is the covariance function for input environments (s \& t). The essence of this method is that the kernel captures the interdependence between the source and target environments. 
\subsection{Non-Transfer Learning Performance Optimization}
\label{sect:nair}
A performance optimization model with no transfer was proposed by Nair \etal~\cite{nair2017using} in FSE '17. It works as follows: 
\be
\item Sample a small set of measurements of configurations from the target environment.
\item Construct performance model with regression trees. 
\item Predict for near-optimal configurations. 
\ee
The key distinction here is that unlike transfer learners, that use a \textit{different source environment} to build to predict for near-optimal configurations in a target environment, a non-transfer method such as this uses configurations \textit{from within the target} environment to predict for near-optimal configurations.

%% file: other_pseudo.tex
\begin{figure*}[t!]
    \centering
    \includegraphics[width=\linewidth]{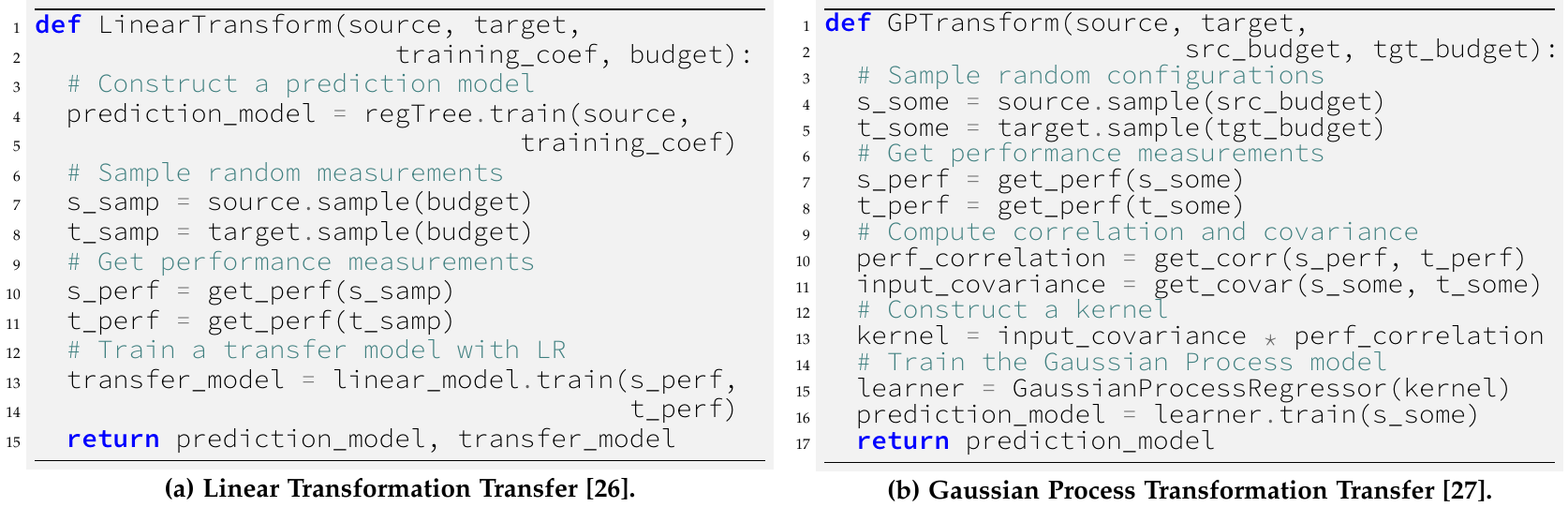}
    \caption{Pseudocodes of other transfer learning methods.}
\end{figure*}
    

%% file: body/6-experimental_setup.tex
\section{Experimental Setup}
\label{sect:expt}
\subsection{Subject Systems}
\label{sect:datasets}
\input{datasets.tex}

In this study, we selected five configurable software systems from different domains, with different functionalities, and written in different programming languages. We selected these real-world software systems since their characteristics cover a broad spectrum of scenarios. Briefly,
\bi
\item {\sc Spear} is an industrial strength bit-vector arithmetic decision procedure and Boolean satisfiability (SAT) solver. It is designed for proving software verification conditions, and it is used for bug hunting. It consists of a binary configuration space with 14 options with $2^{14}$ or 16384 configurations. We measured how long it takes to solve an SAT problem in all $2^{14}$ configurations in 10 environments.

\item {\sc x264} is a video encoder that compresses video files and has 16 configurations options to adjust output quality, encoder types, and encoding heuristics. Due to the cost of sampling the entire configuration space, we randomly sample 4000 configurations  in 21 environments. 

\item {\sc SQLite} is a lightweight relational database management system, which has 14 configuration options to change indexing and features for size compression. Due to the cost of sampling and a limited budget, we use 1000 randomly selected configurations in 15 different environments.

\item {\sc SaC} is a compiler for high-performance computing. The SaC compiler implements a large number of high-level and low-level optimizations to tune programs for efficient parallel executions. It has 50 configuration options to control optimization options. We measure the execution time of the program for 846 configurations in 5 enviroments. 

\item {\sc Storm} is a distributed stream processing framework which is used for data analytics. We measure the latency of the benchmark in 2,048 randomly selected configurations in 4 environments. 
\ei

Table \ref{tab:datasets} lists the details of the software systems used in this paper. Here, $|N|$ is the number of configuration options available in the software system. If the options for each configuration is \textit{binary}, then there can be as much as $2^{|N|}$ possible configurations for a given system\footnote{On the other hand, if there are $|o|$ possible options, then there may be $|o|^N$ possible configurations.}, since it is not possible for us measure the performance of all possible configurations, we measure the performance of a subset of the  $2^{|N|}$ samples, this subset is denoted by $|C|$. The performance of each of the $|C|$ configurations are measured under different hardware ($H$), workloads ($W$), and software versions ($V$). A unique combination of $H, W, V$ constitutes an enviroment which is denoted by $E$. Note that, measuring the performance of $|C|$ configurations in each of the $|E|$ enviroments can be very costly and time consuming. Therefore, instead of all combinations of $H \times W \times V$, we measure the performance in only a subset of the enviroments (the total number is denoted by $|E|$).

\subsection{~Evaluation Criterion}
\label{sect:nar}


Typically, performance models are evaluated based on accuracy or error using measures such as \textit{Mean Magnitude of Relative error} (abbrv. $\mathit{MMRE}$) which is given by:
\[
  MMRE = \frac{|predicted-actual|}{actual} \cdot 100
\]
It has recently been shown that exact measures like MMRE can be somewhat misleading to assess configurations~\cite{foss2003simulation, nair2017using, nair2018finding}. An alternative is to use \textit{rank-based metrics} that compute the difference between the {\em relative rankings} of the performance scores~\cite{nair2017using, nair2018finding}.  The key intuition behind {\em relative rankings} is that the raw accuracy (as measured by MMRE) is less important than the rank-ordering of configurations from best to worst. As long as a model can preserve the order of the rankings of the configurations, it is still possible to determine which configuration is the most optimum. We can quantify this by measuring the differences in ranking between the actual rank and the predicted rank. More formally, rank-difference $R^\delta$ is measured as:
\[
R^\delta = |\mathit{Rank(Predicted)} - \mathit{Rank(Actual)}|
\]
We note that rank difference is still not particularly informative. This is because it ignores the distribution of performance scores and \textit{a small difference in performance measure can lead to a large rank difference and vice-versa}~\cite{trubiani2018performance}. 

To illustrate the challenges with $R^\delta$ and $MMRE$, consider the example in \cref{fig:nar_eg} where we are trying to find a configuration with the \textit{minimum} value. Here, although the difference between the predicted value and the actual value is only $0.02$, the rank difference $R^\delta$ is $90$.  But this does not tell us if $R^\delta = 90$ is good or bad. While, in the same~\cref{fig:nar_eg}, when we calculate MMRE we get an error of only $22\%$, this may convey a false sense of accuracy. In the same example, let us say that the maximum value permissible is $0.11$, then according to~\cref{fig:nar_eg}, our predicted value for the best performance (which recall is supposed to the lowest) is the \textit{highest permissible value} of $0.11$. 

\begin{figure}[t!]
    \centering
    \small
    \begin{tabular}{l|l|l}
        & Value & Rank\\\hline
    Actual & 0.09 & 100\\\hline
    Predicted & 0.11 & 10\\\hline
    Difference & 0.02 & 90
    \end{tabular}
    
    \[
    MMRE = \frac{0.11-0.09}{0.09} \times 100 = 22\%
    \]
    \[
    R^\delta = |10-100| = 90
    \]
    Now, let's say the $min=0.09$ and $max=0.11$. Then, 
    \[
    NAR = \frac{0.11 - 0.09}{0.11-0.09} \times 100 = 100\%
    \]
    \caption{A contrived example to illustrate the challenges with MMRE and rank based measures}
    \label{fig:nar_eg}
\end{figure}


\input{RQ1.tex}



Therefore, to obtain a realistic estimate of optimality of a configuration, in this paper, we propose a measure called \textit{Normalized Absolute Residual} (NAR) inspired by \textit{Generational Distance} or \textit{Inverted Generation Distance} used commonly is search based software engineering~\cite{wang2016practical, chen2018sampling, deb2002fast}. It represents the ratio of (a) difference between the actual performance value of the optimal configuration and the predicted performace value of the optimal configuration, and (b) The absolute difference between the \textit{maximum} and \textit{minimum} possible performace values. Formally: 
\begin{equation}
    \setlength{\belowdisplayskip}{0.2em}
    \setlength{\abovedisplayskip}{0.2em}
\label{eq:nar}
{ \tiny
  \mathit{NAR} = \frac{|min(f(c)) - f(c^{*})|}{max(f(c)) - min(f(c))} \cdot 100
}
\end{equation}
Where $min(f(c))$ is the value of the true minima of configuration $c$, $f(c^{*})$ is the predicted value of the minima, and $max(f(c))$ is the largest performance value of a configuration. This measure is equivalent to Absolute Residual between predicted and actual, normalized to lie between 0\% to 100\% (hence the name \textit{Normalized Absolute Residual} or $NAR$). According to this formulation, the \textit{lower the $NAR$}, the better. Reflecting back on~\cref{fig:nar_eg}, we see that the $NAR$ is 100\% which is exact what is expected when a predicted ``minima'' ($0.11$) is equal to the actual ``maxima'' (also $0.11$).


\subsection{Statistical Validation}
\label{sect:stats}

Our experiments are all subjected to inherent randomness introduced by sampling configurations or by a different source and target environments. To overcome this, we use 30 repeated runs, each time with a different random number seed. The repeated runs provide us with sufficiently large sample size for statistical comparisons. Each repeated run collects the values of NAR.

To rank these 30 numbers collected as above, we use the Scott-Knott test 
recommended by Mittas and Angelis~\cite{mittas13}:
\bi
\item
A list of treatments, sorted by their mean value, are split at the point that maximizes the expected value of the difference in their mean before after the split.
\item
That split is accepted if, between the two splits, (a) there is  
a statistically significant difference using a hypothesis test $\mathcal{H}$, and (b) the difference between the two splits is {\em not} due to a small effect.
\item 
Recurse on both splits if the split is acceptable.
\item
Once no more splits are found, they are ``ranked'' smallest to largest (based on their median value).
\ei
In our work, in order to judge the statistical significance we use a non-parametric bootstrap test with 95\% confidence~\cite{efron93}. Also, to make sure that the statistical significance is not due to the presence of small effects, we use an A12 test~\cite{Vargha00}. Briefly, the A12 test measures the probability that one split has a lower NAR values than another. If the two splits are equivalent, then $A12 = 0.5$. Likewise if $A12\geq0.6$, then 60\% of the times, values of one split are significantly smaller that the other. In such a case, it can be claimed that there is a \textit{significant effect} to justify the hypothesis test. We use these two tests (bootstrap and A12) since these are non-parametric and have been previously demonstrated to be informative~\cite{leech2002call, poulding10, arcuri11, shepperd12a, kampenes07, Kocaguneli2013:ep}.

%% file: datasets.tex
\begin{table*}[tbp!]

\begin{center}
\resizebox{\linewidth}{!}{%
\begin{tabular}{l|p{1cm}|p{1.55cm}|p{2.5cm}|p{3cm}|p{1.5cm}|p{3cm}@{}}
\hline
\rowcolor{lightgray}\textbf{System} & \textbf{Language} & \textbf{\{$|C|$, $N$, $|E|$\}} & \multicolumn{1}{c|}{\textbf{$H$}} & \multicolumn{1}{c|}{\textbf{$W$}} & \multicolumn{1}{c|}{\textbf{$V$}} & \multicolumn{1}{c}{\textbf{Unchanged}}\bigstrut\\\hline

\rowcolor[HTML]{FFFFFF}
x264 & C, Assembly & 4000, 16, 21 & NUC/4/1.30/15/SSD NUC/2/2.13/7/SSD Station/2/2.8/3/SCSI, AWS/1/2.4/1.0/SSD, AWS/1/2.4/0.5/SSD, Azure/1/2.4/3/SCSI & 8/2, 32/11, 128/44 & r2389, r2744 & Memory, CPU, background services \bigstrut\\\hline
SPEAR & C, Assembly & 16384, 14, 10 & NUC/4/1.30/15/SSD, NUC/2/2.13/7/SSD, Station/2/2.8/3/SCSI, AWS/1/2.4/1.0/SSD, AWS/1/2.4/0.5/SSD, Azure/1/2.4/3/SCSI & (in \#variables/\#clauses), 774/5934, 1008/7728, 1554/11914, 978/7498 & 1.2, 2.7 & Memory, CPU, background services\bigstrut\\\hline
\rowcolor[HTML]{FFFFFF}
SQLite & C & 1000, 14, 15 & NUC/4/1.30/15/SSD, NUC/2/2.13/7/SSD, Station/2/2.8/3/SCSI, AWS/1/2.4/1.0/SSD, AWS/1/2.4/0.5/SSD, Azure/1/2.4/3/SCSI & write--seq, read--batc, read--rand, read--seq & 3.7. 6.3, 3.19.0.0 & Memory, CPU, background services\bigstrut\\\hline
SaC & C & 846, 50, 5 & NUC/4/1.30/15/SSD, NUC/2/2.13/7/SSD, Station/2/2.8/3/SCSI, AWS/1/2.4/1.0/SSD, AWS/1/2.4/0.5/SSD, Azure/1/2.4/3/SCSI & random matrix generator, particle filtering, differential, equation solver, k-means, optimal matching, nbody, simulation, conjugate, gradient, garbage collector. & 1.0.0 & Memory, CPU, background services\bigstrut\\\hline
\rowcolor[HTML]{FFFFFF}
Storm & Clojure & 2048, 12, 4 & NUC/4/1.30/15/SSD, NUC/2/2.13/7/SSD, Station/2/2.8/3/SCSI, AWS/1/2.4/1.0/SSD, AWS/1/2.4/0.5/SSD, Azure/1/2.4/3/SCSI & WordCount, RollingCount, RollingSort, SOL & Storm 0.9.5 + Zookeeper 3.4.11 & JVM machine, Zookeeper Options, Memory, CPU, background services\bigstrut\\\hline
\end{tabular}
}
\caption{{\small Overview of the real-world subject systems. $|C|$:Number of Configurations sampled per environment, $N$=Number of configuration options, $|E|$: Number of Environments, $|H|$: Hardware, $|W|$: Workloads, and $|V|$: Versions.}}
\label{tab:datasets}
\end{center}
\end{table*}

%% file: RQ1.tex
\begin{figure*}[t]
{\scriptsize
    \begin{minipage}[]{0.475\linewidth}
    \textbf{{\sc X264}}\vspace{1mm}\\
    \resizebox{\linewidth}{!}{%
    \arrayrulecolor{black}%
    \begin{tabular}{|l|l|r|r|c|}
    \arrayrulecolor{black}
    \hline\rowcolor{lightgray}{\small \textbf{Rank}} & {\small\textbf{Dataset}} & {\small \textbf{Median}} & {\small\textbf{IQR}} & \bigstrut[t]\\\hline  
    \rowcolor{lightergray}  1 &      x264\_18 &    0.35  &  1.82 & \quart{0}{2}{0}{1} \\
    \rowcolor{lightergray}  1 &       x264\_9 &    0.35  &  1.62 & \quart{0}{2}{0}{1} \\
    \hline  2 &      x264\_10 &    0.94  &  8.25 & \quart{0}{12}{1}{1} \\
      2 &       x264\_7 &    0.94  &  8.25 & \quart{0}{12}{1}{1} \\
      2 &      x264\_11 &    1.62  &  7.46 & \quart{1}{11}{2}{1} \\
     3 &      x264\_16 &    2.33  &  12.18 & \quart{0}{18}{3}{1} \\
      3 &       x264\_2 &    2.33  &  12.18 & \quart{0}{18}{3}{1} \\
    3 &       x264\_6 &    2.82  &  5.35 & \quart{0}{8}{4}{1} \\  
    3 &      x264\_20 &    3.65  &  13.74 & \quart{1}{20}{5}{1} \\
      4 &      x264\_19 &    6.95  &  41.97 & \quart{3}{62}{10}{1} \\
        4 &      x264\_17 &    13.61  &  32.32 & \quart{5}{48}{20}{1} \\
      4 &      x264\_13 &    16.42  &  51.65 & \quart{2}{76}{24}{1} \\
      4 &      x264\_15 &    20.14  &  50.68 & \quart{4}{75}{29}{1} \\
      5 &      x264\_14 &    27.24  &  42.74 & \quart{3}{63}{40}{1} \\
      5 &       x264\_0 &    28.63  &  49.77 & \quart{6}{73}{42}{1} \\
    \hline \end{tabular}}\\[0.5mm]
    \textbf{{\sc Sac}}\vspace{1mm}\\
    \resizebox{\linewidth}{!}{%
    \begin{tabular}{|l|l|r|r|c|}
    \arrayrulecolor{black}
    \hline\rowcolor{lightgray}{\small \textbf{Rank}} & {\small\textbf{Dataset}} & {\small \textbf{Median}} & {\small\textbf{IQR}} & \\\hline
    \rowcolor{lightergray}  1 &        sac\_6 &    0.27  &  0.14 & \quart{0}{0}{0}{0} \\
    \hline  2 &        sac\_4 &    0.96  &  4.26 & \quart{0}{3}{0}{0} \\
      2 &        sac\_8 &    1.04  &  3.67 & \quart{0}{3}{0}{0} \\
      2 &        sac\_9 &    2.29  &  4.98 & \quart{1}{4}{1}{0} \\
      3 &        sac\_5 &    10.8  &  89.65 & \quart{8}{71}{8}{0} \\
    \hline \end{tabular}}\\
    \textbf{{\small {\sc Storm} }}\vspace{1mm}\\
    \resizebox{\linewidth}{!}{%
    \begin{tabular}{|l|l|r|r|c|}
    \arrayrulecolor{black}
    \hline\rowcolor{lightgray}{\small \textbf{Rank}} & {\small\textbf{Dataset}} & {\small \textbf{Median}} & {\small\textbf{IQR}} & \bigstrut[t]\\\hline  
    \rowcolor{lightergray}  1 & storm\_feature9 &    0.0  &  0.0 & \quart{0}{0}{0}{1999} \\
    \rowcolor{lightergray}  1 & storm\_feature8 &    0.0  &  0.0 & \quart{0}{0}{0}{1999} \\
    \rowcolor{lightergray}  1 & storm\_feature6 &    0.0  &  0.01 & \quart{0}{19}{0}{1999} \\
    \rowcolor{lightergray}  1 & storm\_feature7 &    0.01  &  0.04 & \quart{0}{79}{19}{1999} \\
    \hline \end{tabular}}
    \end{minipage}\hspace{10pt}
    \begin{minipage}[]{0.475\linewidth}
    \textbf{{\small {\sc Spear}}}\\[0.1cm]
    \resizebox{\linewidth}{!}{%
    \begin{tabular}{|l|l|r|r|c|}
    \arrayrulecolor{black}
    \hline\rowcolor{lightgray}{\small \textbf{Rank}} & {\small\textbf{Dataset}} & {\small \textbf{Median}} & {\small\textbf{IQR}} & \bigstrut[t]\\\hline  
    \rowcolor{lightergray} 1 &      spear\_7 &    0.1  &  0.1 & \quart{0}{1}{1}{13} \\
    \rowcolor{lightergray}  1 &      spear\_6 &    0.1  &  0.2 & \quart{0}{2}{1}{13} \\
    \rowcolor{lightergray}  1 &      spear\_1 &    0.1  &  0.1 & \quart{0}{1}{1}{13} \\
    \rowcolor{lightergray}  1 &      spear\_9 &    0.1  &  0.5 & \quart{0}{6}{1}{13} \\
    \rowcolor{lightergray}  1 &      spear\_8 &    0.1  &  0.2 & \quart{0}{2}{1}{13} \\
    \rowcolor{lightergray}  1 &      spear\_0 &    0.1  &  0.91 & \quart{0}{12}{1}{13} \\
    \hline  2 &      spear\_5 &    0.28  &  0.3 & \quart{2}{4}{3}{13} \\
      3 &      spear\_4 &    0.6  &  1.17 & \quart{5}{16}{8}{13} \\
      4 &      spear\_2 &    1.09  &  5.31 & \quart{8}{71}{14}{13} \\
      5 &      spear\_3 &    1.89  &  4.48 & \quart{18}{61}{25}{13} \\
    \hline \end{tabular}}\\[0.2cm]
    \textbf{{\small {\sc Sqlite}}}\\[0.1cm]
    \resizebox{\linewidth}{!}{%
    \begin{tabular}{|l|l|r|r|c|}
    \arrayrulecolor{black}
    \hline\rowcolor{lightgray}{\small \textbf{Rank}} & {\small\textbf{Dataset}} & {\small \textbf{Median}} & {\small\textbf{IQR}} & \bigstrut[t]\\\hline  
    \rowcolor{lightergray}  1 &    sqlite\_17 &    0.8  &  1.13 & \quart{0}{1}{0}{0} \\
    \rowcolor{lightergray}  1 &    sqlite\_59 &    2.0  &  3.44 & \quart{0}{4}{2}{0} \\
    \rowcolor{lightergray}  1 &    sqlite\_19 &    2.0  &  4.88 & \quart{0}{6}{2}{0} \\
    \hline  2 &    sqlite\_44 &    1.96  &  6.91 & \quart{0}{10}{2}{0} \\
      2 &    sqlite\_16 &    2.52  &  7.41 & \quart{1}{10}{3}{0} \\
      2 &    sqlite\_73 &    2.82  &  7.24 & \quart{1}{10}{3}{0} \\
      2 &    sqlite\_45 &    3.47  &  11.86 & \quart{0}{17}{4}{0} \\
      2 &    sqlite\_10 &    3.88  &  6.92 & \quart{1}{9}{4}{0} \\
      2 &    sqlite\_96 &    4.94  &  6.04 & \quart{1}{8}{6}{0} \\
      2 &    sqlite\_79 &    5.64  &  5.24 & \quart{4}{7}{7}{0} \\
      2 &    sqlite\_11 &    6.64  &  5.75 & \quart{5}{8}{8}{0} \\
      2 &    sqlite\_52 &    6.84  &  7.95 & \quart{1}{11}{8}{0} \\
      2 &    sqlite\_97 &    7.68  &  13.71 & \quart{7}{18}{10}{0} \\
      3 &    sqlite\_18 &    13.17  &  54.68 & \quart{5}{74}{17}{0} \\
      3 &    sqlite\_94 &    27.43  &  47.66 & \quart{9}{65}{37}{0} \\
    \hline \end{tabular}}
    \end{minipage}
    \caption{ {\small Median NAR of 30 repeats. Median NAR is the normalized absolute residual values  as described in Equation~\ref{eq:nar}, and IQR the difference between 75th percentile and 25th percentile found during multiple repeats. Lines with a dot in the middle (~\protect\quartex{-2}{13}{6}), show the median as a round dot within the IQR. All the results are sorted by the median NAR: a lower median value is better. The left-hand column (\textit{Rank}) ranks the various techniques where lower ranks are better. Overall, we find that there is always at least one environment, denoted in \colorbox{lightergray}{light gray}, that is much superior (lower NAR) to others.}}
    \label{fig:rq1}
}

\end{figure*}

%% file: body/7-research_questions.tex
\section{Results}
\label{sect:results}
\input{RQ2.tex}
\noindent\textbf{\textsf{RQ1: Does there exist a Bellwether Environment?}}\label{subsec:rq1}

\noindent\textbf{\textit{\underline{Purpose:}}} The first research question seeks to establish the presence of bellwether environments within different environments of a software system. If there exists a bellwether environment, then identifying that environment can greatly reduce the cost of finding a near-optimal configuration for different environments.\\
\textbf{\textit{\underline{Approach:}}} For each subject software system, we use the environments to perform a pair-wise comparison as follows:
\be
\item We pick one environment as a source and evaluate all configurations to construct a regression tree model.
\item The remaining environments are used as targets. For every target environment, we use the regression tree model constructed above to predict for the best configuration.
\item Then, we measure the NAR of the predictions (see \S\ref{sect:nar}). 
\item Afterwards, we repeat steps 1, 2, and 3 for all the other source environments and gather the outcomes.
\ee
We repeat the whole process above 30 times and use the Scott-Knott test to rank each environment best to worst.

\noindent\textbf{\textit{\underline{Result:}}}
Our results are shown in Fig.~\ref{fig:rq1}.  \begin{steelblue}
Overall, we find that there is always at least one environment (the bellwether environment) in all the subject systems, that is much superior to others. Note that, {\sc Storm} is an interesting case, where all the environments are ranked 1, which means that all the environments are equally useful as a bellwether environment
---in such cases, any randomly selected environment could serve as a bellwether. 
Further, we note that the variance in the bellwether environments are much lower compared to other environments. Low variance indicates the low median NAR is not an effect of randomness in our experiments and hence increases our confidence in the existence of bellwethers.\end{steelblue}

Please note, in this specific experiment, we use \textit{all} measured configurations (i.e., 100\% of $|C|$ in~\cref{tab:datasets}) to determine if bellwethers exist. This ensures that the existence of bellwethers is not biased by how we sampled the configuration space. Later, in RQ2, we will restrict our study to determine what fraction of the samples would be adequate to find the bellwethers.

\begin{steelblue}One may be tempted to argue that the answer to this question trivially could be answered as "yes" since it is unlikely that all environments exhibit identical performance and there will always be some environment that can make better predictions. However, observe that the environments ranked first performs much better than the rest (with certain exceptions), and hence, the difference between the bellwether environment and others is not coincidental. Further, by exhaustively comparing the performance of all available environments, we demonstrate that it is ill advised to randomly pick any available source lest we risk choosing a sub-optimal configuration setting.\end{steelblue} 

\medskip
\begin{result}
    In each subject system, there exist bellwether environment(s) which can be used to find the near-optimal configurations for the rest of the environments.
\end{result}

\begin{table}[t]
\centering
\setlength{\abovecaptionskip}{0.5\baselineskip}
\setlength{\belowcaptionskip}{\baselineskip} {
\arrayrulecolor{black}
\caption{{\small  Effectiveness of source selection method.}}
\label{tbl:method}
\resizebox{\linewidth}{!}{%
\begin{tabular}{@{}crrrrrr@{}}
\toprule
\multicolumn{1}{c}{\multirow{2}{*}{Subject System}} & \multicolumn{2}{l}{100\% Samples} & \multicolumn{2}{l}{FindBellwether} & \multicolumn{2}{l}{Difference ($\Delta_\%$)} \\ \cmidrule(l){2-7} 
\multicolumn{1}{c}{} & Median & IQR & Median & IQR & Median & IQR \\ \midrule
SQLite & 0.8 & 1.13 & 1.8 & 2.48 & 1.0 & 1.35 \\
Spear & 0.1 & 0.1 & 0.1 & 0 & 0.0 & 0.0 \\
x264 & 0.35 & 1.62 & 0.9 & 1.06 & 0.55 & 0.16 \\
Storm & 0.0 & 0.0 & 0.0 & 0.0 & 0.0 & 0.0 \\
SaC & 0.27 & 0.14 & 0.63 & 7.4 & 0.36 & 6.9 \\ \bottomrule
\end{tabular}}}
\vspace{-2em}
\end{table}

\newpage
\noindent\textbf{\textsf{RQ2: How many measurements are required to discover bellwether environments?}}

\noindent\textbf{\textit{\underline{Purpose:}}} The bellwether environments found in RQ1 required us to use 100\% of the measured performance values from all the environments\footnote{Note, except for SPEAR, we only have measured a subset of all possible configuration space since we were limited by the time and the cost required to make exhaustive measurements}. Sampling all configurations may not be practical, since that may take an extremely long time~\cite{jamshidi2017transfer2}. 
Thus, we ask if we can find the bellwether environments sooner using fewer samples. Further, we ask how many such samples are required.\\
\textbf{\textit{\underline{Approach:}}}
\begin{steelblue}
We used the racing algorithm discussed in Section~\tion{finding} to incrementally sample the configurations until a bellwether environment has been discovered. It works as follows:
    
\be
    \item We start from 1\% of configurations from each environment and assume that every environment is a potential bellwether environment.
    \item Then, we increment the number of configurations in steps of 1\% and measure the NAR values. 
    \item We rank the environments and eliminate those that do not show much promise. 
    \item We repeat the above steps until we cannot eliminate any more environments.
\ee

When the above \textit{discover} process terminates, we note that only a fraction of the available samples are used to discover the bellwether. We measure the number of samples required for estimating the bellwether. Further, to understand if the smaller sample size is sufficient to identify a near-optimal configuration, we compare the performance of the discovered bellwether environment with 100\% with the predicted bellwether environment using a smaller sample size.
\end{steelblue}

\noindent
\begin{steelblue}
\noindent\textbf{\textit{\underline{Result:}}}
Table~\ref{tbl:method} summarizes our findings. We find: 
\begin{itemize}
  \item In all 5 cases, the racing algorithm for finding bellwether terminated after using the following percentage of samples:
   \be
    \item \textit{x264}: $10.21\%$ of 4000 samples
    \item \textit{SQLite}: $11.42\%$ of 1000 samples
    \item \textit{Spear}: $13.79\%$ of 16384 samples
    \item \textit{SaC}: $15.4\%$ of 846 samples
    \item \textit{Storm}: $17.40\%$ of 2048 samples
  \ee
  \item Further, from Table~\ref{tbl:method}, when compared with the NAR values obtained with using all 100\% of the available samples (Columns 2 and 3) to the NAR values when using only the fraction required to find the bellwether using the racing algorithm (Columns 4 and 5), we see that the difference which is formally is given by $\Delta_\% = \left| \mathit{NAR}_{100\%} - \mathit{NAR}_{10\%}\right|$ is very minimal. We note that these differences ($Delta_\%$) are:
\be
\item 1\% in \textit{SQLite};
\item 0\% in \textit{Spear} and \textit{Storm};
\item 0.55\% in \textit{x264}; and
\item 0.36\% in \textit{SaC}
\ee
\end{itemize}
These results are most encouraging in that we need only about 10\% of the samples to determine the bellwether:
\end{steelblue}

\begin{result}
The bellwether environment can be recognized using only a fraction of the measurements (under 10\%). Encouragingly, the identified bellwether environments have similar NAR values to the bellwether environment with 100\% of samples.
\end{result}

\noindent\textbf{\textsf{RQ3: How does \tool compare with other non-transfer-learning based methods?}}
\label{sect:rq2}

\begin{figure}[!t]
  \centering
  \includegraphics[width=\linewidth]{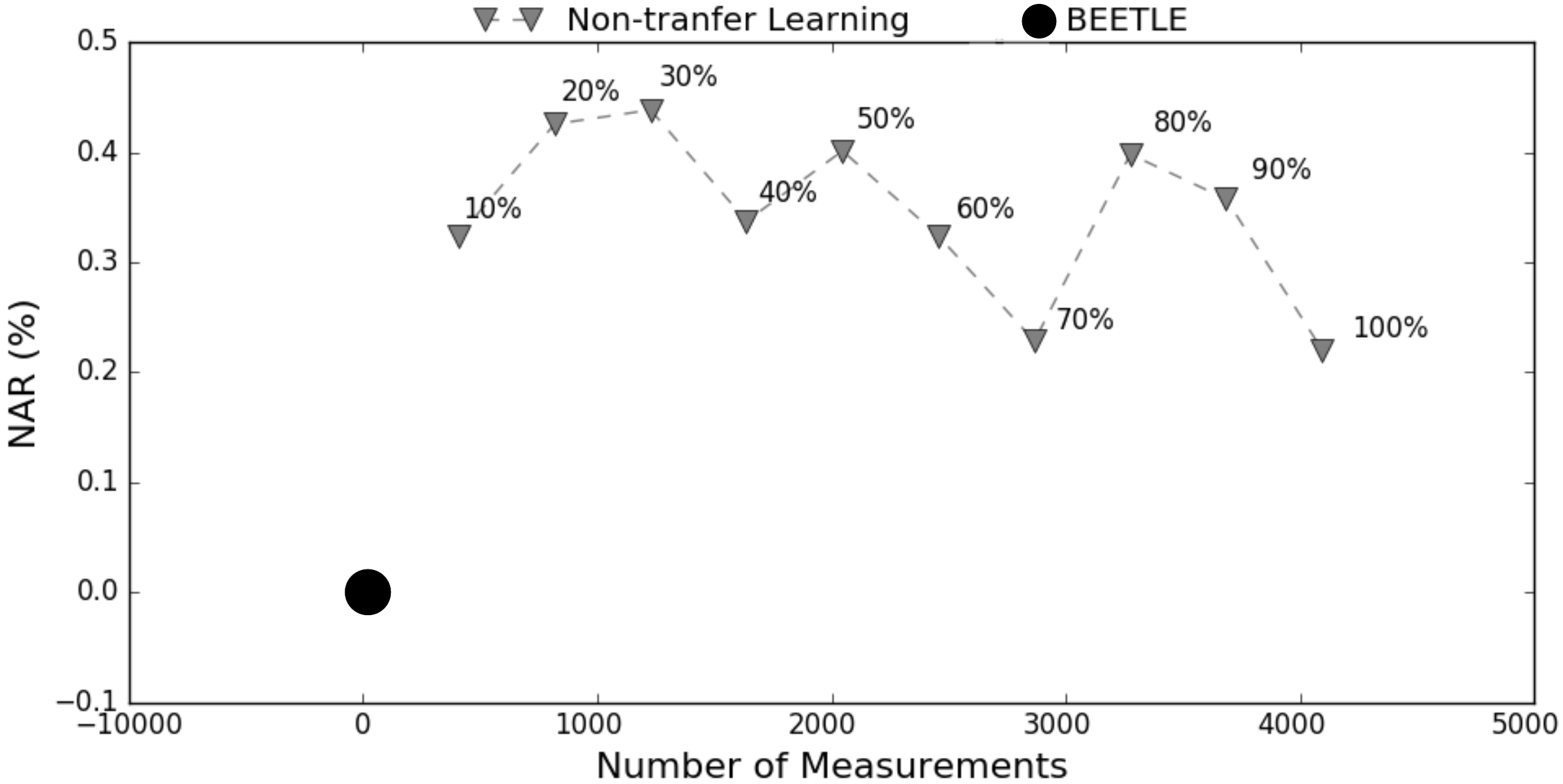}
  \caption{{\small Trade-off between the quality of the configurations and the cost to build the model for {\sc x264}. The cost to find a good configuration using bellwethers is much lower than that of non-transfer-learning methods. }}
  \label{fig:tradeoffx264}
\end{figure}

\noindent\textbf{\textit{\underline{Purpose}}}: We explore how \tool compares to a non-transfer learning approach. For our experiment, we use the non-transfer performance optimizer proposed by Nair \etal~\cite{nair2017using}.
 \begin{steelblue}Both \tool and Nair \etal's methods seek to achieve the same goal---find the optimal configuration in a target environment. \tool uses configurations from a \textit{different source} to achieve this, whereas the non-transfer learner uses configurations from \textit{within the target}. Please note \tool can use anywhere between 0\%--100\% of the configurations from the bellwether environment. In the previous RQs, we showed that 10\% was adequate when using the bellwether environment.\end{steelblue}\\
\textbf{\textit{\underline{Approach:}}} Our setup involves evaluating the Win/Loss ratio of \tool to the non-transfer learning algorithm while predicting for the optimal configuration. Comparing against true optima, we define ``win'' as cases where \tool has a better (or same) $NAR$ as the non-transfer learner. If the non-transfer learner has a better $NAR$, that counts as a ``loss''.\\
\textbf{\textit{\underline{Result:}}}
Our results are shown in~\cref{fig:rq2_1,fig:tradeoffx264}. In \cref{fig:rq2_1}, the x-axis represents the number of configurations (expressed in \%) to train the non-transfer learner and \tool, and the y-axis represents the number of wins/losses. We observe:
\begin{itemize}
\item \textit{Better performance:} In $\frac{4}{5}$ systems, \tool ``wins'' significantly more than it ``losses''. This means that \tool is better than (or similar to) non-transfer learning methods. 
\item \textit{Lower cost:} Regarding cost, we note that \tool outperforms the non-transfer learner significantly, ``winning'' at configurations of 10\% to 100\% of the original sample size. Further, when we look at the trade-off between performance and number of measurements in \cref{fig:tradeoffx264}, we note that \tool achieves a NAR close to zero with around 100 samples. Also, the non-transfer learning method of Nair \etal~\cite{nair2017using} has significantly larger NAR while also requiring more samples.
\end{itemize}
 
\begin{result}
     \tool performs better than (or same as) a non-transfer learning approach. \tool is also cost/time efficient as it requires far fewer measurements.
  \end{result}

\noindent\textbf{\textsf{RQ4: How does \tool compare to state-of-the-art methods?}}\label{subsec:rq4}

\begin{figure}[tbp]
  \centering
  \includegraphics[width=0.85\linewidth]{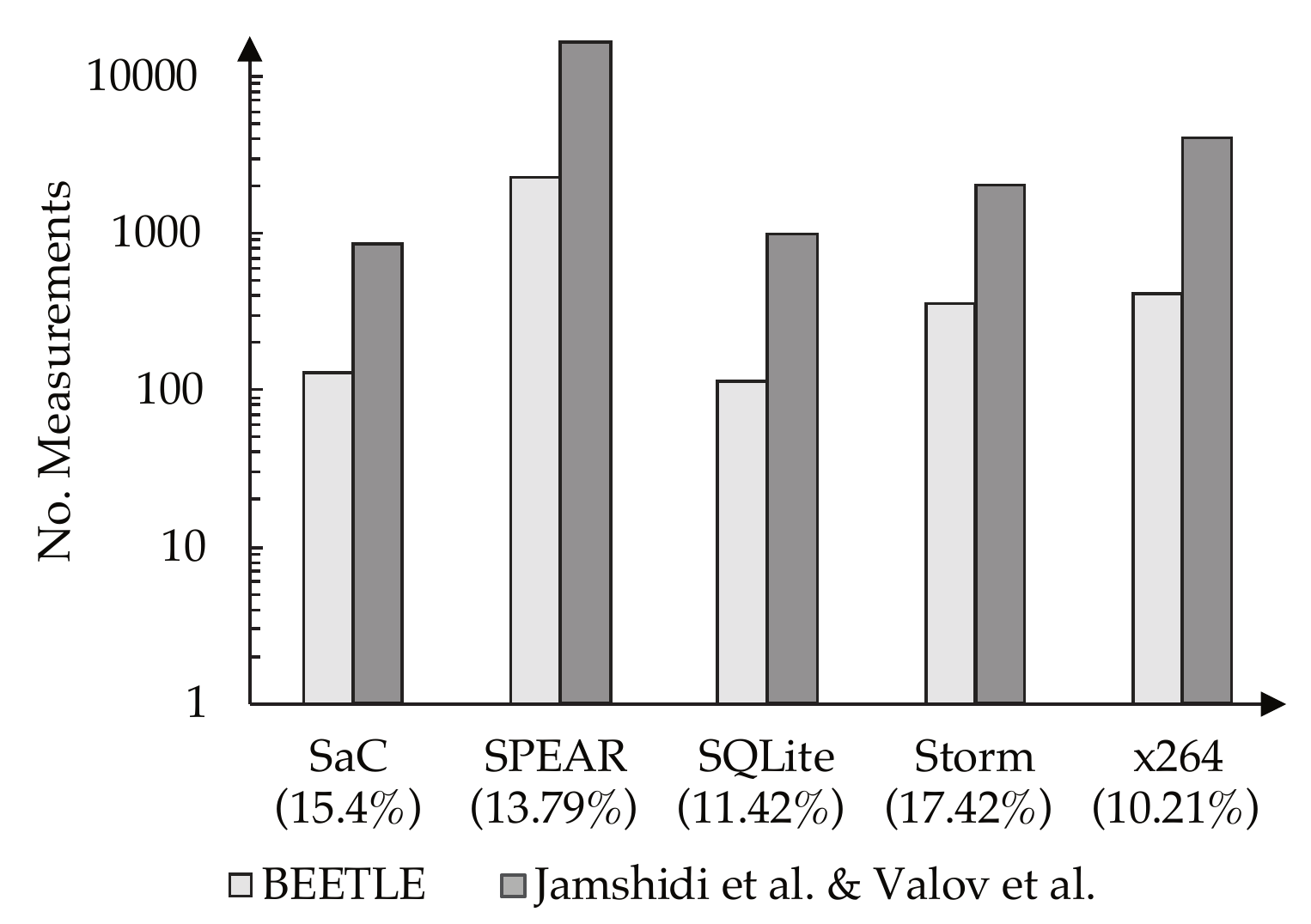}
  \caption{{\small  Number of samples required by \tool (in \colorbox{lightergray}{light gary}) v/s the other two state-of-the-art transfer learners (in \colorbox{lightgray}{gray}). Note: The other two transfer learners require that \textit{all available data} be used for transfer learning therefore the chart shows one bar for both transfer learners.}}
  \label{fig:rq4_measurement}
\end{figure}

\noindent\textbf{\textit{\underline{Purpose:}}} The main motivation of this work is to show that the source environment can have a significant impact on transfer learning. In this research question, we seek to compare \tool with other state-of-the-art transfer learners by Jamshidi \etal~\cite{jamshidi2017transfer} and Valov \etal~\cite{valov2017transferring}.\\
\textbf{\textit{\underline{Approach:}}} We perform transfer learning the methods proposed by Valov \etal~\cite{valov2017transferring} and Jamshidi \etal~\cite{jamshidi2017transfer} (see~\tion{tl}). Then we measure the NAR values and compare them statistically using Skott-Knott tests. Finally, we rank the methods from best to worst based on their Skott-Knott ranks.\\
\textbf{\textit{\underline{Result:}}}
Our results are shown in Fig.~\ref{fig:rq4}. In this figure, the best transfer learner is ranked 1. We note that in 4 out of 5 cases, \tool performs just as well as (or better than) the state-of-the-art. This result is encouraging in that it points to a significant impact on choosing a good source environment can have on the performance of transfer learners. Further, in~\cref{fig:rq4_measurement} we compare the number of performance measurements required to construct the transfer learners (note the logarithmic scale on the vertical axis). Here, we note that \tool uses an order of magnitude fewer samples ($\approx$13\% on average) that the other methods. The total number of available samples for each software system is shown in the second column of~\cref{tab:datasets}~(see values corresponding to $|C|$). 
Based on these results, we note that \tool requires far fewer measurements compared to the other transfer-learning methods. That is,

\begin{result}
        \tool performs just as well as (or better than) other state-of-the-art transfer learners for performance optimization using far fewer measurements.
\end{result}

\begin{figure}[!t]
{\scriptsize
  \begin{minipage}[]{\linewidth}
    \arrayrulecolor{black}
  \textbf{{\sc Sac}}\\[0.05cm] 
  \resizebox{\linewidth}{!}{%
  \begin{tabular}{|l|l|r|r|c|}
  \arrayrulecolor{black}
  \hline\rowcolor{lightgray}{\small \textbf{Rank}} & {\small\textbf{Learner}} & {\small \textbf{Median}} & {\small\textbf{IQR}} & \\\hline 
  \rowcolor{lightergray} 1 &    Jamshidi \etal~\cite{jamshidi2017transfer} &  1.58 & 5.39 & \quart{0}{4}{0}{0} \\\hline
   2 &   \tool &  6.89 & 99.1 & \quart{0}{79}{5}{0} \\
   2 &   Valov \etal~\cite{valov2017transferring} &  6.99 & 99.24 & \quart{0}{79}{6}{0} \\
  \hline \end{tabular}}\\[0.01cm]
  \textbf{{\sc Spear}}\\[0.05cm]
  \resizebox{\linewidth}{!}{%
  \begin{tabular}{|l|l|r|r|c|}
  \arrayrulecolor{black}
  \hline\rowcolor{lightgray}{\small \textbf{Rank}} & {\small\textbf{Learner}} & {\small \textbf{Median}} & {\small\textbf{IQR}} & \\\hline 
  \rowcolor{lightergray} 1 &    Jamshidi \etal~\cite{jamshidi2017transfer} &  0.70 & 1.29 & \quart{0}{0}{3}{0}\\
  \rowcolor{lightergray}1 &   \tool &  0.79 & 1.40 & \quart{0}{0}{3}{0} \\
  \rowcolor{lightergray}1 &   Valov \etal~\cite{valov2017transferring} &  1.11 & 1.98 & \quart{0}{2}{4}{0} \\
  \hline\end{tabular}}\\[0.01cm]
  \textbf{{\sc SQLite}}\\[0.05cm]
  \resizebox{\linewidth}{!}{
  \begin{tabular}{|l|l|r|r|c|}
  \arrayrulecolor{black}
  \hline\rowcolor{lightgray}{\small \textbf{Rank}} & {\small\textbf{Learner}} & {\small \textbf{Median}} & {\small\textbf{IQR}} & \\\hline 
  \rowcolor{lightergray} 1 &   \tool &  5.41 & 9.28 & \quart{2}{10}{5}{0} \\\hline
   2 &   Valov \etal~\cite{valov2017transferring} &  6.96 & 12.91 & \quart{3}{15}{6}{0} \\
   3 &   Jamshidi \etal~\cite{jamshidi2017transfer}  &  18.51 & 50.85 & \quart{2}{50}{18}{0} \\
  \hline \end{tabular}}\\[0.01cm]
  \textbf{{\sc Storm}}\\[0.05cm]
  \resizebox{\linewidth}{!}{%
  \begin{tabular}{|l|l|r|r|c|}
  
  \arrayrulecolor{black}
  \hline\rowcolor{lightgray}{\small \textbf{Rank}} & {\small\textbf{Learner}} & {\small \textbf{Median}} & {\small\textbf{IQR}} & \\\hline 
  \rowcolor{lightergray}  1 &   \tool &  0.04 & 0.06 & \quart{0}{0}{0}{0} \\\hline
   1 &    Jamshidi \etal~\cite{jamshidi2017transfer} &  0.86 & 20.69 & \quart{0}{21}{1}{0} \\
   2 &   Valov \etal~\cite{valov2017transferring} &  2.47 & 53.98 & \quart{0}{54}{4}{0} \\
  \hline \end{tabular}}\\[0.01cm]
  \textbf{{\sc x264}}\\[0.05cm]
  \resizebox{\linewidth}{!}{%
  \begin{tabular}{|l|l|r|r|c|}
  \arrayrulecolor{black}
  \hline\rowcolor{lightgray}{\small \textbf{Rank}} & {\small\textbf{Learner}} & {\small \textbf{Median}} & {\small\textbf{IQR}} & \\\hline 
  \rowcolor{lightergray} 1 &   \tool &  8.67 & 27.01 & \quart{2}{31}{8}{0} \\\hline
   2 &   Valov \etal~\cite{valov2017transferring} &  16.99 & 41.24 & \quart{5}{47}{17}{0} \\
   3 &    Jamshidi \etal~\cite{jamshidi2017transfer} &  43.58 & 28.39 & \quart{34}{48}{44}{0} \\
  \hline \end{tabular}}
  \end{minipage}}
  \caption{{\small Comparison between state-of-the-art transfer learners and \tool. The best transfer learner is shaded \colorbox{lightergray}{gray}.
   The ``ranks'' shown in the left-hand-side column come from the statistical analysis described in \tion{stats}.}}
  \label{fig:rq4}
  \end{figure}

%% file: RQ2.tex
\begin{figure*}[t]

	\centering
\begin{subfigure}[t]{0.195\linewidth}
		\centering
		\includegraphics[width=\linewidth]{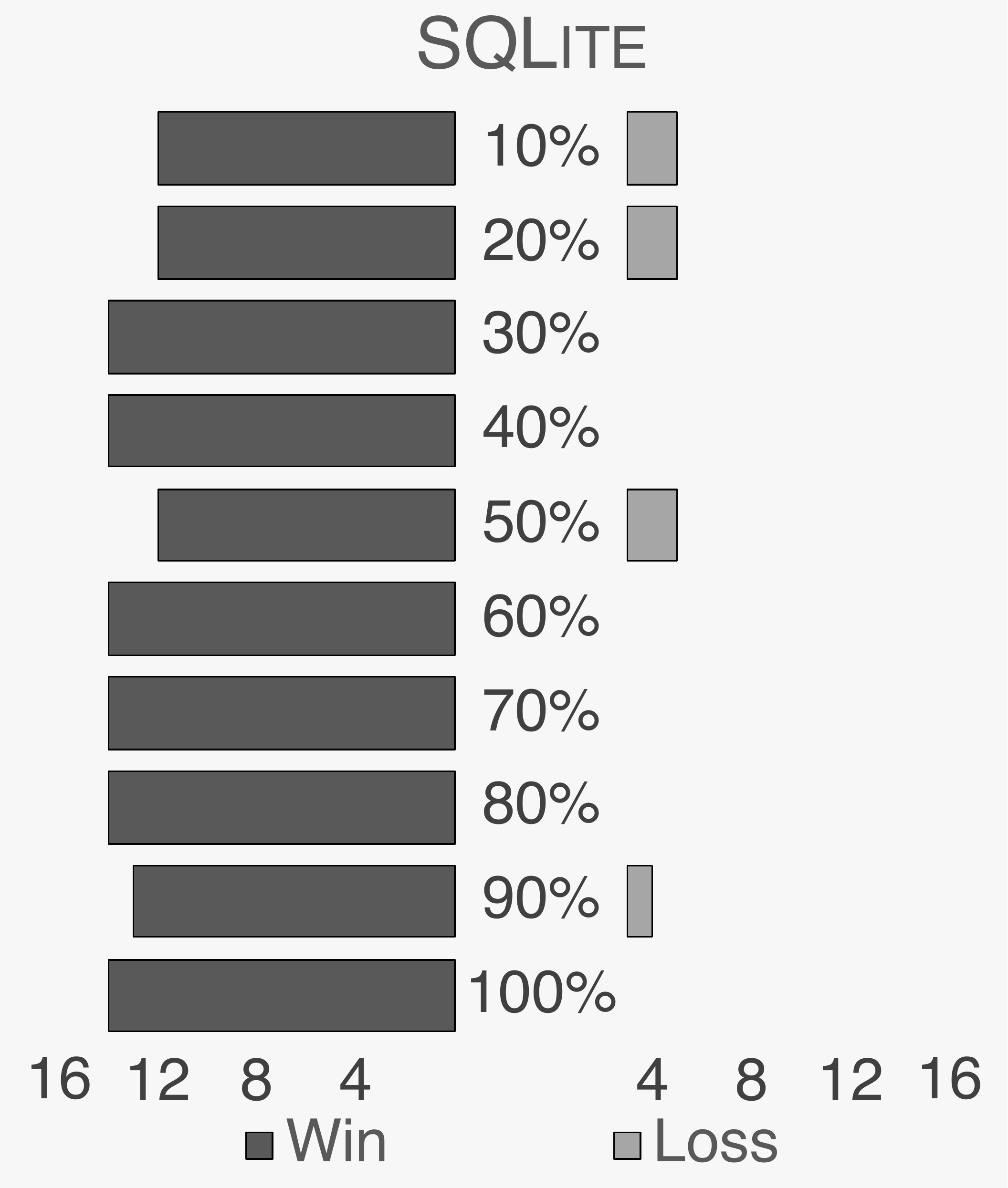}
\end{subfigure}~%
\begin{subfigure}[t]{0.195\linewidth}
		\centering
		\includegraphics[width=\linewidth]{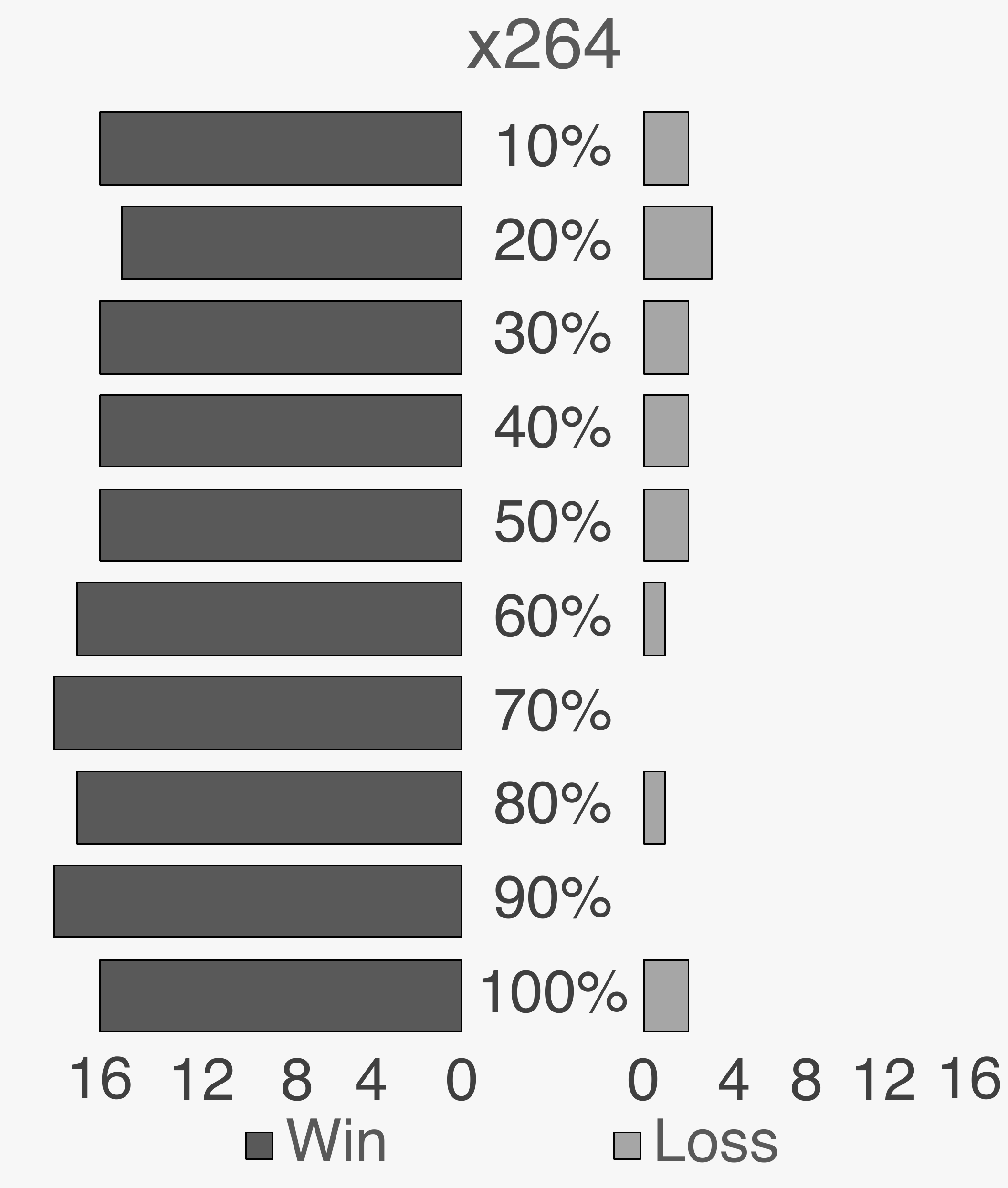}
\end{subfigure}~%
\begin{subfigure}[t]{0.195\linewidth}
		\centering
		\includegraphics[width=\linewidth]{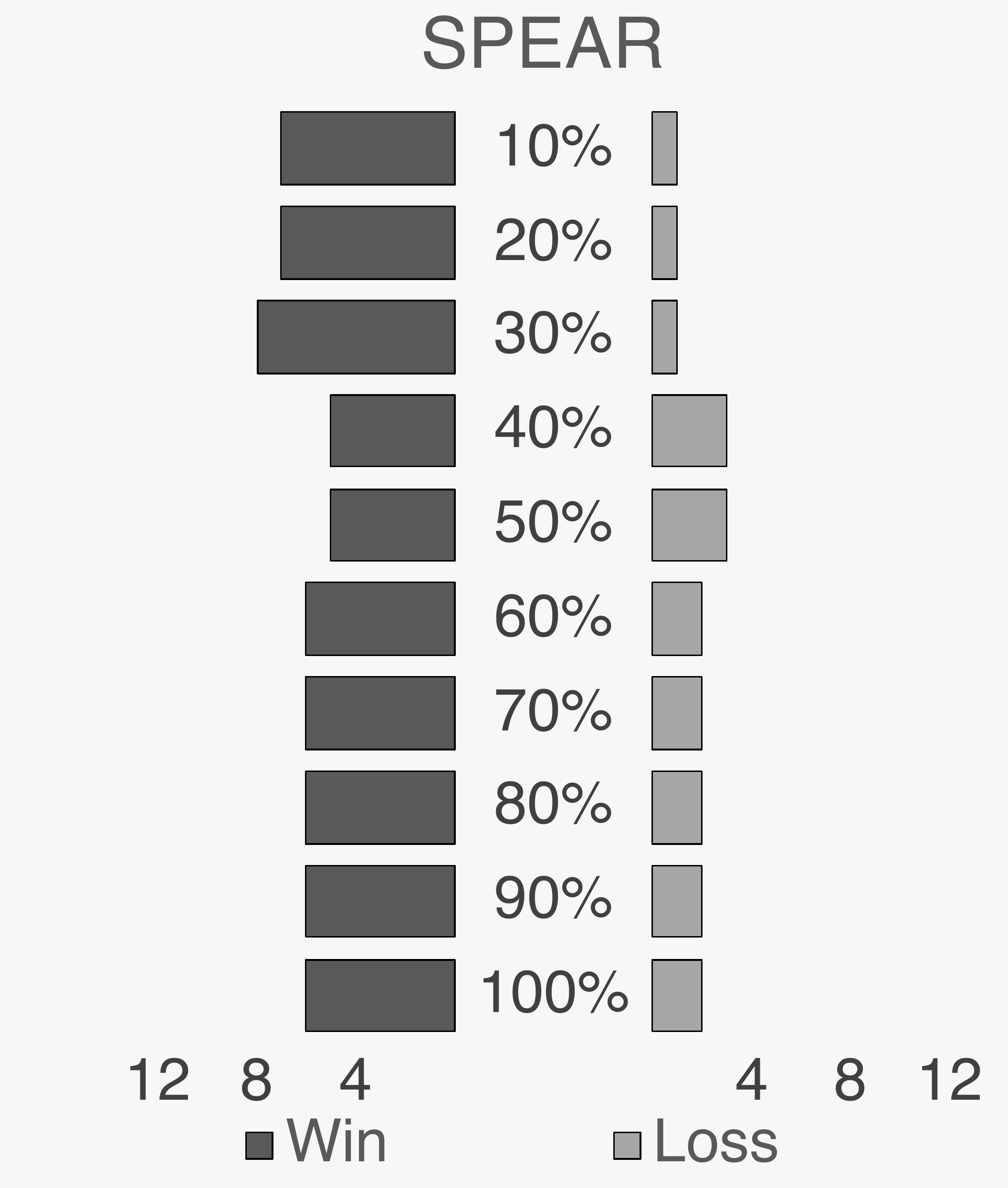}
\end{subfigure}~%
\begin{subfigure}[t]{0.195\linewidth}
		\centering
		\includegraphics[width=\linewidth]{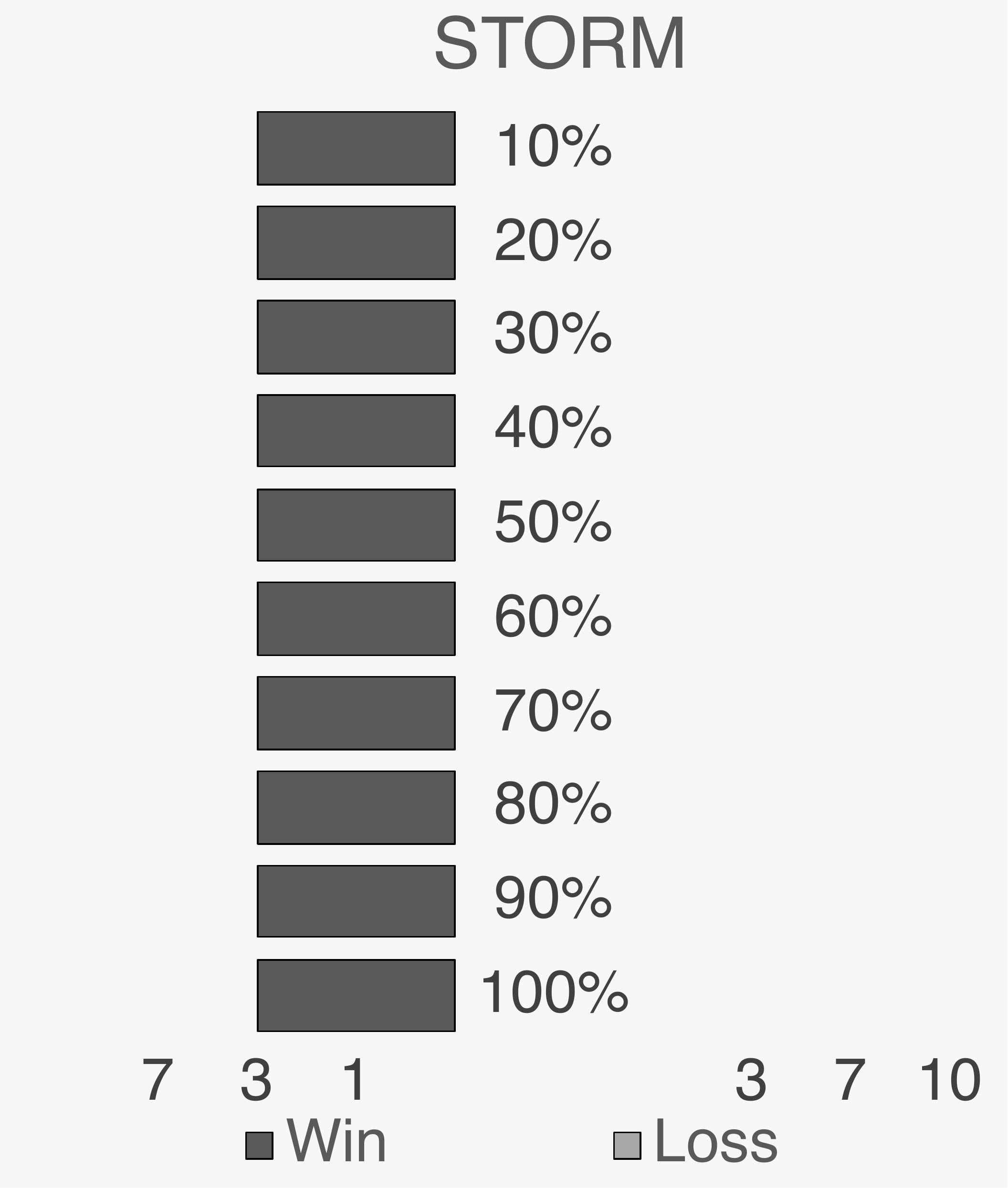}
\end{subfigure}~%
\begin{subfigure}[t]{0.195\linewidth}
		\centering
		\includegraphics[width=\linewidth]{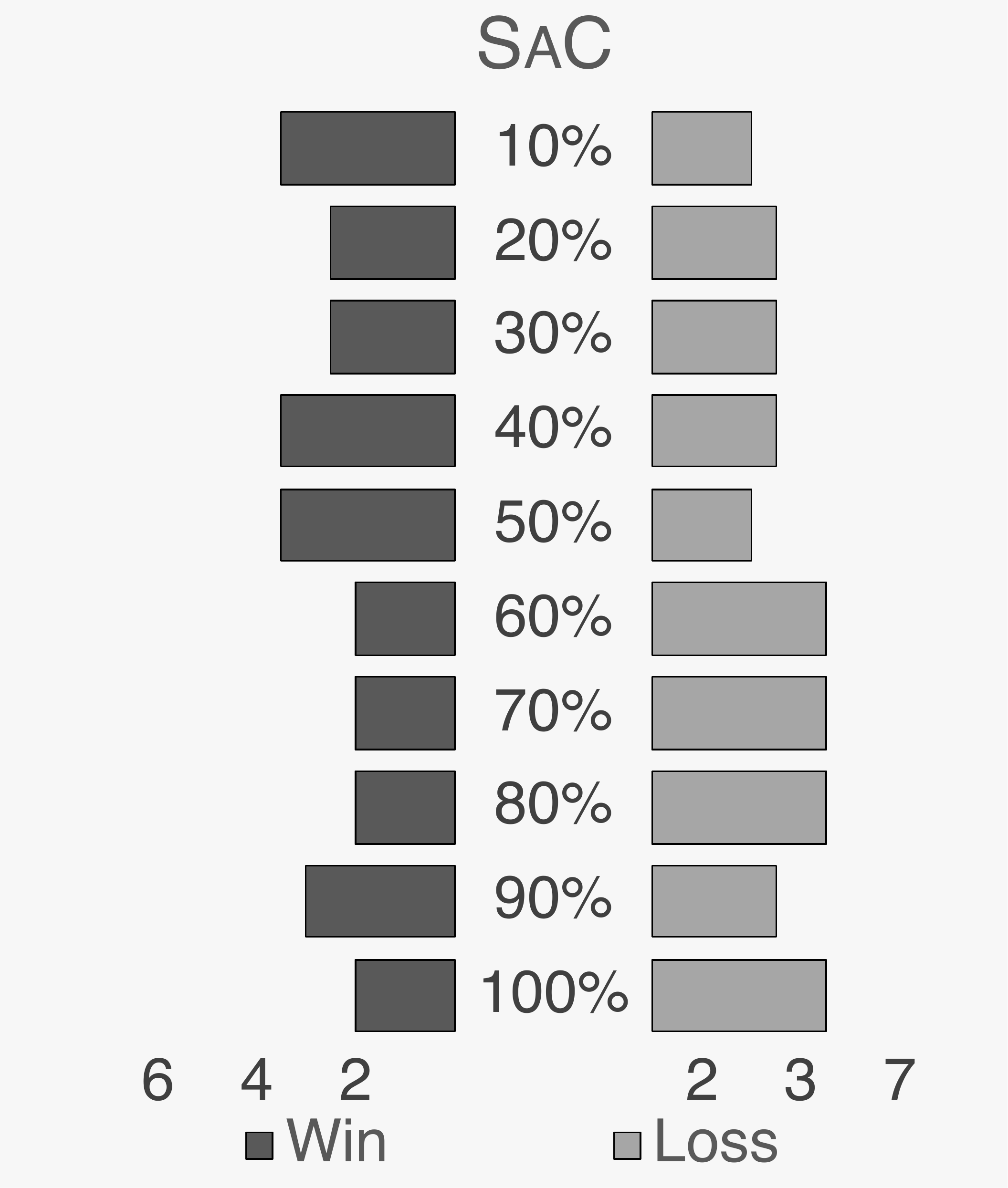}
\end{subfigure}
	\caption{{\small ~Win/Loss analysis of learning from the bellwether environment and target environment using Scott Knott. The x-axis represents the \% of available samples used to build a model. The y-axis is the count.
  }}
	\label{fig:rq2_1}
\end{figure*}

%% file: body/8-discussion.tex
\section{Discussion}
\label{sect:disc}

This section addresses some additional questions that may arise with regards to \tool's real-world applicability.

\noindent\textbf{What is the effect of \tool on the day to day business of a software engineer?}
From an industrial perspective, \tool can be used in at least the following ways:
\begin{itemize}
\item
Consider an organization which has to optimize their software system for different clients (who have different workload and hardware---different AWS subscriptions). While on-boarding new clients, the company might not be able to afford to invest extensive resources in finding the near-optimal configuration to appease the client. State-of-the-art transfer learning techniques would expect the organization to provide a source workload (or environment) for this task. But without a subject matter expert (SME) with the relevant knowledge, it is hard for humans to select a suitable source. \textit{\tool removes the need for such SMEs since it automates \underline{source selection}, along with transferring knowledge between the source and the target environment}.
\item
Consider an organization, which needs to migrate all their workload from a legacy platform to a different cloud platform (e.g., AWS to AZURE or vice versa). Such an organization now has many workloads that they need to optimize; however, they lack experience and performance measurements, on the new platform to accomplish this goal. \textit{In such cases, \tool provides a way to discover an ideal \underline{source} to transfer knowledge to enable efficient migration of workloads}. 
\end{itemize}

\noindent\textbf{How complex is \tool compared to other methods?}
\tool is among the easiest transfer learning methods currently available. In comparison with the state-of-the-art methods studied here, we require only few measurements of software systems running under different environments, we can build a $\mathit{find_bellwether}$ method that comprises of an all-pairs round-robin comparison followed by elimination of poorly performing enviroments. Then, transfer learning uses one of many off-the-shelf machine learners to build a prediction model (here we use Regression Trees). In this paper, we demonstrate that this method is just as powerful as other methods while being an order of magnitude cheaper in terms of the number of measurements required. 

\begin{figure}[!b]
    \centering
    \includegraphics[width=\linewidth]{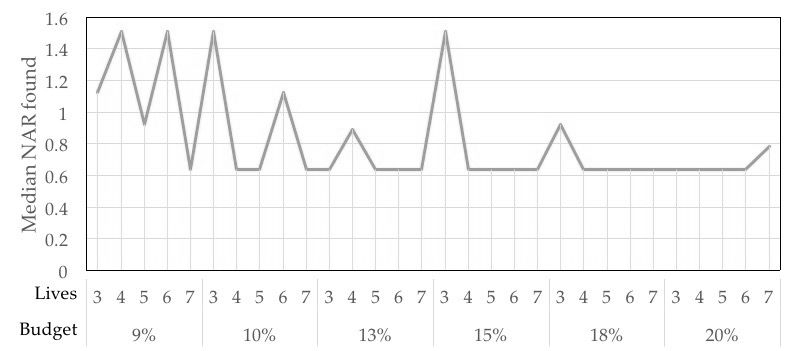}
    \caption{ The trade-off between the budget of the search, the number of lives, and the NAR (quality) of the solutions for x264. Performance depends on the budget and number of lives, i.e., as the budget increases the NAR value decreases; likewise, as the number lives increases, the NAR improves. }
    \label{fig:tuning}
  \end{figure} 

\noindent{\bfseries What are the impact of different hyperparameter choices?} With all the transfer learners and predictors discussed here, there are a number of internal parameters that may (or may not) have a significant impact on the outcomes of this study. We identify two key hyperparameters that affect \tool namely, \textit{Budget} and \textit{Lives}. As shown in \Cref{fig:approach_b}(d), both these hyperparameters determine when to stop sampling the source and declare the bellwethers. These bellwethers subsequently affect transfer learning. To study the effect of these hyperparameters, we plot the trade-off between the budget and lives versus NAR. This is shown in~\cref{fig:tuning}. Here,
\begin{itemize}
    \item \textit{Budget:} There is discernible impact of larger budget on the performance of bellwethers.  We note that the performance is directly related to the budget, i.e., as the budget increases the NAR value decreases (lower NAR values are better). This is to be expected, an increased budget permits an larger sample to construct transfer learners, thereby improving the likelihood of finding a near optimal solution.
    \item \textit{Lives:} Although lower lives seems to correspond to larger NAR (worse). The relationship between the number of lives and NAR is less pronounced than that between Budget and NAR. That said, we noted that having 5 or lives generally corresponds to better NAR values. Thus, in all the experiments in this paper, we use 5 lives as default.
\end{itemize}

\noindent\textbf{Is \tool applicable in other domains?} In principle, yes. \tool could be applied to any transfer learning application, where the choice of the source data impacts the performance of transfer learning. This can be applied to problems such as configuring big data systems~\cite{JC:MASCOTS16}, finding suitable cloud configuration for a workload~\cite{Hsu2018scout, hsu2017low}, 
configuring hyperparameters of machine learning algorithms~\cite{fu2016tuning, afridi2018}, runtime adaptation of robotic systems~\cite{jamshidi2017transfer}. In these applications, the correct choice of source datasets using bellwethers can help to reduce the amount of time it takes to discover a near-optimal configuration setting.

\noindent\textbf{Can \tool identify bellwethers in completely dissimilar environments?}
In theory, yes. Given a software system, \tool currently looks for an environment which can be used to find a near-optimal configuration for a majority of other environments for \textit{that} software system. Therefore, given performance measurements in various environments, \tool can assist in discovering a suitable source environment to transfer knowledge across environments comprised of different hardware, software versions, and workloads.

\noindent\textbf{When are bellwethers ineffective?}
The existence of bellwethers depends on the following: 
\begin{itemize}
\item
\textit{Metrics used:} Finding bellwether using metrics that are not justifiable, may be unsuccessful, for example, discovering bellwethers in performance optimization, by measuring MMRE instead of NAR may fail~\cite{nair2017using}.
\item
\textit{Different Software System:} Bellwethers of a certain software system `A' may not work for software system `B.' In other words, it cannot be used for cases where the configuration spaces across environment are not consistent.
\item
 \textit{Different Performance Measures: } Bellwether discovered for one performance measure (time) may not work for other performance measures (throughput).
\end{itemize}

%% file: body/9-threats.tex
\section{~Threats To Validity}
\label{sect:threats}
\begin{steelblue}
As with any empirical study, biases can affect the final results. Therefore, any conclusions of this work
must be considered with the following issues in mind:
\bi
\item \textit{Evaluation Bias}: 
In  {\bf  RQ2, RQ3} and {\bf RQ4}, we have shown the performance of \tool by comparing them using statistical tests on their \measure to draw conclusions regarding their performance when compared to other transfer learning and non-transfer-learning learning methods. While those results are true, the conclusions are scoped by the evaluation metrics we used to write this paper (i.e., \measure). It is possible that with other measurements, there may be slightly different conclusions. This is to be explored in future research.

\item \textit{Construct Validity}: At various places in this report, we made engineering decisions about (e.g.) choice of machine learning models (in our case decision tree regression), step-size for incremental sampling, etc. While these decisions were made using advice from the literature, we acknowledge that other constructs might lead to other conclusions. 

\item \textit{External Validity}: For this study, we have selected a diverse set of subject systems, and a large number of environment changes from the data collected by Jamshidi \etal~\cite{jamshidi2017transfer2} for their studies. The performance measures were gathered on known software environments such as AWS, Azure, and NUC. There is a possibility that measurement of other performance measures or availability of additional performance measures may result in a different outcome. Therefore, one has to be careful when generalizing our findings to other subject systems and environment changes. Even though we tried to run our experiment on a variety of software systems from different domains, we do not claim that our results generalize beyond the specific case studies explored here. That said, to enable reproducibility, we have shared our scripts and the gather performance data.  

\item \textit{Statistical Validity}: To increase the validity of our results, we applied Scott-Knott tests (which in turn comprises of two statistical tests, bootstrap, and the a12). Hence, anytime in this paper, we reported that ``X was different from Y'', then that report was based on Scott-Knott tests.
 
\item \textit{Sampling Bias}: Our conclusions are based on the performance measure of the five software systems collected by Jamshidi \etal~\cite{jamshidi2017transfer2} for their studies. Different initial samples may have lead to different conclusions. That said, we note that our samples are sufficiently large, so we have some confidence that these samples represent an interesting range of configurations and their performances. As evidenced by our results that are remarkably stable over 30 repeated runs.

\item \textit{Learner Bias}: There are various models used in performance optimization such as Gaussian Process~\cite{jamshidi2017transfer}, Regression Trees~\cite{guo2013variability, sarkar2015cost, nair2018finding}, and Bagging, Random Forest, and Support Vector Machines (SVMs)~\cite{valov2015empirical}. It is possible that changing the learner used may change our findings. However, we strive to minimize the uncertainty by choosing Decision Tree Regressor, which is the machine learning algorithm that has most consistently been used in the domain of performance modeling and optimization~\cite{guo2013variability, sarkar2015cost, nair2018finding}. Further, we have made available our replication package that enables one to replace Decision Tree with any other machine learning model quickly. 
\ei
\end{steelblue}

%% file: body/10-related_work.tex
\section{Related Work}
\label{sect:related}


\noindent\textbf{Performance Optimization: }Modern software systems come with a large number of configuration options. 
For example, in {\sc Apache} (a popular web server) there are around 600 
different configuration options and in {\sc hadoop}, as of version 2.0.0, there are around 150 different configuration options, and the number of options is constantly growing~\cite{xu2015hey}. These configuration options control the 
internal 
properties of the system such as memory and response times. Given the large number of configurations, it becomes increasingly difficult to assess the impact of the configuration options on the system's performance. To address this issue, a common practice is to employ performance prediction models to estimate the performance of the system under these 
configurations~\cite{guo2013variability, hoste2006, hutter2014, thereska2010, 
valov2015, westermann12}. To leverage the full benefit of a software system and its features, researchers augment performance prediction models to enable
\textit{performance 
optimization}~\cite{nair2017using,oh2017finding}.

Performance optimization is an essential challenge in software engineering. As shown in the next few paragraphs, this problem has attracted much recent research interest. Approaches 
that use meta-heuristic 
search algorithms to explore the configuration space of Hadoop for high-performing configurations have been proposed~\cite{tang2018searching}. 
It has been reported that such meta-heuristic search can find configurations options that perform significantly better than baseline default configurations. 
In other work, a
control-theoretic framework called \textit{SmartConf} to automatically set and 
dynamically adjust performance-sensitive configurations to optimize 
configuration options~\cite{wang2018understanding}. For the specific case of 
deep learning clusters, a job scheduler called 
\textit{Optimus} has been developed to determine
configuration options that optimize training speed and resource 
allocations~\cite{peng2018optimus}. Performance optimization has also been extensively explored in other domains 
such as Systems Research~\cite{zhu2017bestconfig, li2018understanding} 
and Cloud Computing~\cite{alipourfard2017cherrypick, yadwadkar2017selecting, 
hsu2017low, Hsu2018scout, hsu2018micky}.

Much of the performance optimization tasks introduced above require access to measurements of the software system under various configuration settings. 
However, obtaining these performance measurements can cost a significant amount
of time and money. For example, in one of the software systems studied here ({\sc x264}), it takes over 1536 hours to obtain performance measurements for 11 out the 16 possible configuration options~\cite{valov2017transferring}. This is in addition to other time-consuming tasks involved in commissioning these systems such as setup, 
tear down, etc. Further, making performance measurements can cost an exorbitant amount of money, e.g, it cost several thousand dollars to obtain of 2048 configurations on {\sc x264} deployed in AWS \texttt{c4.large}.

\noindent\textbf{Transfer Learning: }
When a software system is deployed in a new environment, not every user can 
afford to repeat the costly process of building a new performance model to find 
an optimum configuration for that new environment. Instead, researchers propose 
the use of transfer learning to 
reuse the measurements made for previous 
environments~\cite{valov2017transferring, jamshidi2017transfer, 
chen2011experience,golovin2017google}.
Jamshidi \etal~\cite{jamshidi2017transfer}, conducted a preliminary exploratory study of transfer learning in performance optimization to identify transferable knowledge between a source and a target environment, ranging from easily exploitable relationships to more subtle ones. 
They demonstrated that information about influential configuration options could be exploited in transfer learning and that knowledge about performance
behavior can be transferred.

Following this, a number of transfer learning methods were developed to predict 
for the optimum configurations in a new \textit{target} environment, using the 
performance measures of another \textit{source} environment as a proxy. Several 
researchers have shown that transfer 
learning can decrease the cost of learning 
significantly~\cite{jamshidi2017transfer2,  valov2017transferring, 
jamshidi2017transfer,chen2011experience}.

All transfer learning methods place implicit faith in the quality of the source. A poor source can significantly deteriorate the performance of transfer learners.

\noindent\textbf{Source Selection with Bellwethers: }
It is advised that the source used for transfer learning must be chosen with care to ensure optimum performance~\cite{yosinski2014transferable, 
long2015, afridi2018}. An incorrect choice of the source may result in the all too 
common \textit{negative transfer} 
phenomenon~\cite{afridi2018, ben2003, rosenstein2005,pan2010}. A negative transfer can be particularly damaging in that it often leads to performance 
degradation~\cite{jamshidi2017transfer2, afridi2018}. 
A preferred way to avoid negative transfer is with \textit{source 
selection}. Many methods have been proposed for identifying a suitable source for transfer learning~\cite{krishna18, krishna16, afridi2018}. Of these, source
selection using the bellwether effect is one of the simplest. It has been effective in several domains of software engineering~\cite{krishna18, 
mensah17a, mensah17b}. 

Besides negative transfer, previous approaches suffer from a lack of scalability. For example, Google Visor~\cite{golovin2017google} Jamshidi \etal~\cite{jamshidi2017transfer} rely on a Gaussian process which known to  not scaling to large amounts of data in high dimensional spaces~\cite{rasmussen2004gaussian}
Accordingly, in this work, we introduce the notion of source selection with bellwether effect for transfer learning in performance optimization. With this, 
we develop a Bellwether Transfer Learner called \tool. We show that, for performance optimization, \tool can outperform both non-transfer and the transfer learning methods.

%% file: body/11-conclusion.tex
\section{Conclusion}
\label{sect:conclusion}
Our approach, \tool, exploits the bellwether effect---there are one or more bellwether environments which can be used to find good configurations for the rest of the environments. We also propose a new transfer learning method, called \tool, which exploits this phenomenon. As shown in this paper, \tool can quickly identify the bellwether environments with only a few measurements ($\approx10\%$) and use it to find the near-optimal solutions in the target environments. Further, after extensive experiments with five highly-configurable systems demonstrating, we show that \tool:
\begin{itemize}
\item Identifies  suitable sources to construct transfer learners;
\item Finds near-optimal configurations with only a small number of measurements (an average of $13.5\%\approx \frac{1}{7}^{th}$ of the available number of samples);
\item 
Performs as well as non-transfer learning approaches; and 
\item
Performs as well as state-of-the-art transfer learners.
\end{itemize}
Based on our experiments, we demonstrate our initial problem--``whence to learn?'' is an important question, and,
\begin{quote}
    A \textit{good source} with a simple transfer learner is \textit{better than source agnostic} complex transfer learners.
\end{quote}

%% file: body/12-bio.tex
\begin{IEEEbiography}[{\includegraphics[width=1in,clip,keepaspectratio]{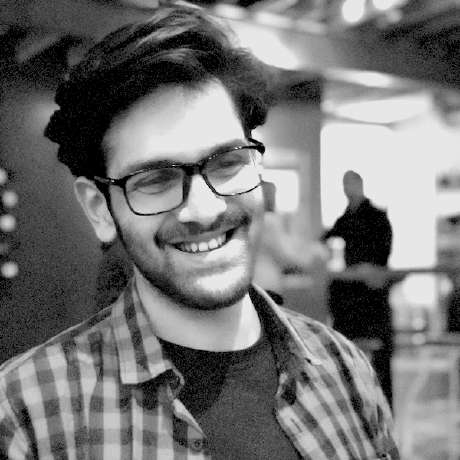}}]{Rahul Krishna} is a post doctoral researcher in Computer Science at Columbia University. He received his Ph.D. in NC State University. His current research explores ways to use machine learning to generate actionable insights for building reliable software systems. His other research interests include program analysis, artificial intelligence, and security. See \url{http://rkrsn.us} for more details.
\end{IEEEbiography}
\vspace{-10mm}
\begin{IEEEbiography}[{\includegraphics[width=1in,clip,keepaspectratio]{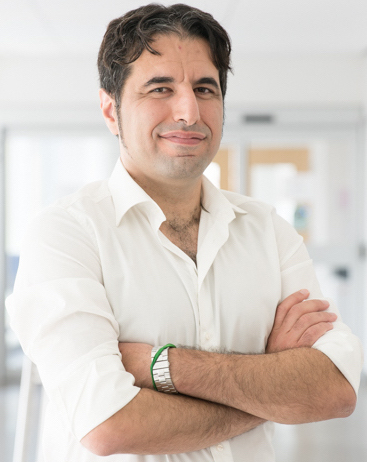}}]{Pooyan Jamshidi} is an Assistant Professor at the University of South Carolina. Pooyan’s general research interests are at the intersection of systems/software and machine learning. He directs the AISys Lab (\url{https://pooyanjamshidi.github.io/AISys/}), where he investigates the development of novel algorithmic and theoretically principled methods for machine learning systems. Prior to his current position, he was a research associate at Carnegie Mellon University and Imperial College London, where he primarily worked on transfer learning for performance understanding of highly-configurable systems.
\end{IEEEbiography}

\begin{IEEEbiography}[{\includegraphics[width=1in,clip,keepaspectratio]{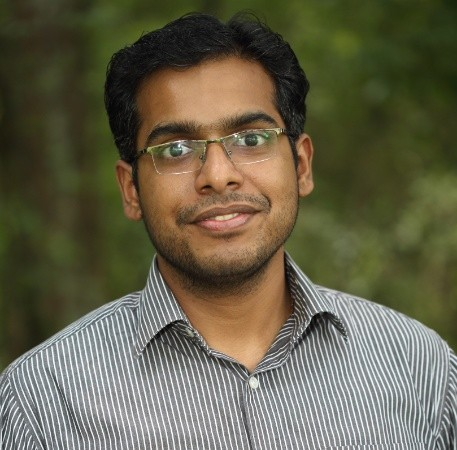}}]{Vivek Nair} graduated with a Ph.D. from the Department of Computer Science at North Carolina State University.  His primary interest lies in exploring possibilities of using multiobjective optimization to solve problems in Software Engineering. At NCSU, he was working on performance prediction models of highly configurable systems. He received his master's degree and worked in the mobile industry for two years before returning to graduate school. For more details, visit \url{http://vivekaxl.com}.
\end{IEEEbiography}
\vspace{-11mm}
\begin{IEEEbiography}[{\includegraphics[width=1in,clip,keepaspectratio]{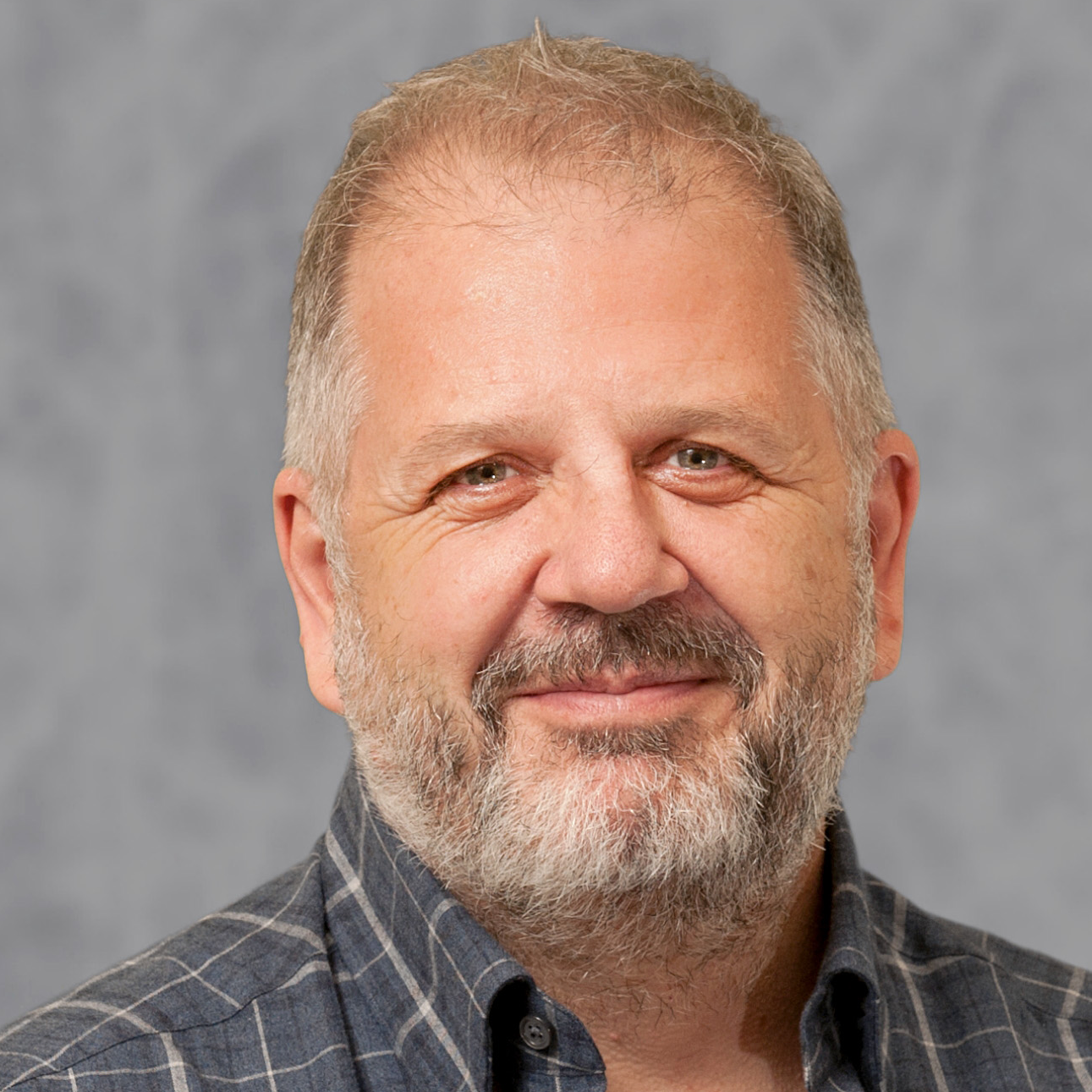}}]{Tim Menzies} (IEEE Fellow)
   is a Professor in CS at NcState.  His research interests include software engineering (SE), data mining, artificial intelligence, search-based SE, and open access science. \url{http://menzies.us}
\end{IEEEbiography}